\documentclass[aps,a4paper,superscriptaddress,twocolumn, nofootinbib]{revtex4} 
\usepackage{tensor}
\usepackage[utf8]{inputenc}
\usepackage{graphicx}
\graphicspath{ {images/} }
\usepackage{amsmath}
\usepackage{amssymb}
\usepackage{enumerate}
\usepackage{subfigure}
\usepackage{tabularx}
\usepackage{lipsum} 
\usepackage{float} 
\usepackage[colorlinks=true, pdfstartview=FitV, linkcolor=blue, citecolor=red, urlcolor=black, breaklinks=true]{hyperref}
\newcommand{\be}{\begin{equation}}
\newcommand{\ee}{\end{equation}}
\newcommand{\ben}{\begin{eqnarray}}
\newcommand{\een}{\end{eqnarray}}
\newcommand{\bes}{\begin{subequations}}
	\newcommand{\ees}{\end{subequations}}
\def\bal#1\eal{\begin{align}#1\end{align}}

\newcommand{\LL}{{\mathcal L}}

\newcommand{\vfi}{\mathrm{\varphi}}

\newcommand{\qt}[1]{``#1''}
\newcommand{\pd}[2]{\ensuremath{\frac{\partial#1}{\partial#2}}}

\newcommand{\pb}[1]{\ensuremath{\partial_{#1}}}
\newcommand{\vf}[1]{\ensuremath{\varphi_{#1}}}

\newcommand{\An}[1]{\ensuremath{A^{\{#1\}}_{\mu}}}

\newcommand{\Aan}[2]{\ensuremath{A^{\{#1\}}_{#2}}}
\newcommand{\Fn}[1]{\ensuremath{F_{\mu\nu}^{\{#1\}}}}
\newcommand{\Bn}[1]{\ensuremath{B_{#1}}}
\newcommand{\Fun}[1]{\ensuremath{F^{\mu\nu\{#1\}}}}
\newcommand{\Ffn}[2]{\ensuremath{F_{#1}^{\{#2\}}}}
\newcommand{\lie}{$\mathcal{G}_{\lilN}$}
\newcommand{\emag}{$\mathrm{U(1)}\ $}
\newcommand{\Vca}[1]{\ensuremath{\mathcal{V}_{#1}}}
\newcommand{\lilN}{\ensuremath{\scriptscriptstyle{N}}}

\begin{document}
	\title{Generalized Maxwell-Higgs vortices in models with enhanced symmetry}
	
\author{D. Bazeia}\affiliation{Departamento de F\'\i sica, Universidade Federal da Para\'\i ba, 58051-970 Jo\~ao Pessoa, PB, Brazil}
\author{M. A. Liao}\affiliation{Departamento de F\'\i sica, Universidade Federal da Para\'\i ba, 58051-970 Jo\~ao Pessoa, PB, Brazil}
\author{M. A. Marques}\affiliation{Departamento de Biotecnologia, Universidade Federal da Para\'iba, 58051-900 Jo\~ao Pessoa, PB, Brazil}\affiliation{Departamento de F\'\i sica, Universidade Federal da Para\'\i ba, 58051-970 Jo\~ao Pessoa, PB, Brazil}
	
	\begin{abstract}
		Topological vortices in relativistic gauge theories in flat three-dimensional spacetime are investigated. We consider the symmetry $\rm{U(1)}\times...\times \rm{U(1)}$, and for each $\rm{U(1)}$ subgroup, a complex scalar field transforming under its action is introduced, as well as generalized permeabilities through which the subsystems are coupled. We investigate in detail the features of static, finite energy solutions within this class of generalized Maxwell-Higgs models, and study the effect of the winding numbers in the magnetic properties of each subsystem. A BPS bound and the related first order equations are introduced for a large class of models. Finally, we present some specific models and solve their equations of motion to find solutions engendering many distinct features in relation to each other and to the standard Nielsen-Olesen vortex.
	\end{abstract}
	
	\maketitle
\section{Introduction}\label{intro}

Existence of vortex lines in type-II superconductors was famously postulated by Abrikosov in an attempt to explain the magnetic properties of these materials~\cite{abrikosov}. This prediction, based on the Ginzburg-Landau theory~\cite{GL}, was confirmed by experiment soon afterwards~\cite{Cribier, Essmann} and rapidly evolved into a central aspect of the theory, as type-II superconductors became more prevalent. These defects were brought into relativistic field theory by Nielsen and Olesen~\cite{Nielsen}, who found vortex lines in a Maxwell-Higgs model with local \rm{U(1)} symmetry. In the static case, these solutions are equivalent to stationary points of the Ginzburg-Landau (GL) free energy functional. A significant advancement was made by Bogomol'nyi~\cite{bogo}, who found that, at the critical coupling that defines the boundary between type-I and II superconductivity, the energy functional is bounded from below by a number proportional to a topological charge. Solutions that saturate this bound are called BPS after Bogomol'nyi, Prasad and Sommerfield~\cite{bogo,ps} and are of importance in the study of topological structures and supersymmetry, see~\cite{Supersym} for example. Topological vortices play an important role in many branches of physics, a few of which are Bose- Einstein theory~\cite{BE}, superfluids~\cite{feynman} and cosmology~\cite{kibble}, as well as condensed matter physics and optics~\cite{soundandlight, Soskin}. Important reviews on vortices and applications can be found in Refs.~\cite{Vilenkin, Pismen, manton, Shifman}.

Interesting generalizations of the Maxwell-Higgs model are achieved by adding extra scalar and gauge degrees of freedom to the Lagrangian density through enhancement of the $\rm{U(1)}$ symmetry. $\rm{U(1)}\times\rm{Z_2}$ and $\rm{U(1)}\times\rm{U(1)}$ symmetries  are of particular importance, as they have been also used to include a hidden sector, which is of interest in the study of dark matter, and to unveil the presence of vortices with internal structures; see, e.g., Refs.~\cite{Long,SHI,SHI2,LongII, Witten,DM,Schapo,ahep,VOR1,VOR2,HORA,CASA, Forgacs} and references therein. Enhanced symmetry of this kind has also found applications in condensed matter physics~\cite{Lukacs}, including the investigation of multicomponent Bose-Einstein condensates~\cite{Catelani} and Type-1.5 superconductivity~\cite{Babaev}. Moreover, in a recent paper~\cite{multilayered}, a generalized permeability associated to the fields of a hidden sector was used to induce interesting modifications in the profile of the fields. In the presence of rotational symmetry, the modifications manifest themselves as ring-like configurations for the magnetic field. 

The present work is motivated by Ref.~\cite{multilayered}, and we will generalize some results from that paper, most notably in allowing for the introduction of more than one extra \emag factor in the gauge group. However, we do not intend to simply extend the results of the preceding paper to the setting of a greater symmetry. Instead, we will conduct a much more detailed investigation and focus on issues that have not been thoroughly addressed before, such as the influence of topological winding numbers, which had been set to unity in our previous investigations, in the observable results of our models. These quantities may be interpreted as a counter of vortices~\cite{Tong}, and some general results from topology imply that they can be identified with the number of zeros of the scalar field, counted with multiplicity~\cite{Bott}. 

A Lagrangian density invariant under $ \mathcal{G}_{\lilN}\equiv\mathrm{U(1)^{\lilN}}$ gauge transformations will be considered in this work. Symmetries of this kind appear quite naturally in quiver theories~\cite{Quiver, QuiverII, QuiverIII}, where each group in the product decomposition of $\mathcal{G}_{\lilN}$ gives rise to a node in a quiver graph. Quivers are a natural feature of supersymmetry and string theory~\cite{QuiverIII}, where they are often discussed in connection with a Bogomol'nyi bound. A mirror symmetry between strongly coupled and quiver theories has been suggested~\cite{mirror}. 

The \emag factors in \lie need not all be related to electromagnetism or hidden sectors from particle physics. Gauge symmetries are often used to conveniently account for extra degrees of freedom and other redundancies that one may wish to keep in their description of a system, and need not relate to any fundamental interaction. Applications of this kind include~\cite{Niemi, Krokhotin, Melnikov, nematic, turbulence} and may be among the factors in the product group considered here. Being unrelated to electromagnetism, the degrees of freedom associated to these \emag factors give rise to fields that transform independently of the gauge transformations of electromagnetism and possibly of each other which, in our formalism, translates to the fact that each scalar field of the theory is charged under only one \emag subgroup. Although some attention has been given in the literature to the $\rm{U(1)}\times\rm{U(1)}$ case, investigation of the higher symmetry groups which may arise if some of these degrees of freedom applications described above occur simultaneously are scarce.  Local \emag symmetry is also important in the context of Bose-Einstein condensates. In a recent paper, it has been suggested that the enhancement of the superconducting properties of graphene in the presence of a Bose-Einstein condensate may be possible \cite{BEG}. Vortex solutions are relevant in investigations concerning superconducting materials as well as  Bose-Einstein condensates, so that, in light of Ref.~\cite{BEG}, models that allow for the coupling of these two types of vortex states may be relevant for future applications.

In the present study, we have structured the investigation as follows: in Sec.~\ref{General} the fields, Lagrangian, and associated equations of motion for a MH-like model with generalized permeability are introduced. Some of the physical aspects of this generalization are discussed, including properties that we argue would be desirable for both the permeabilities and potential. Next, in Sec.~\ref{sym}, we see how the features discussed in the previous section are manifested in the important setting of symmetric solutions. We introduce some new models in Sec.~\ref{models} and solve their equations of motion to see that they present the properties we argued that should be expected from the nature of this generalization. Finally, we discuss our results in Sec.~\ref{end}, and point out some of the many ways in which the results of this paper could be adapted to different applications.

\section{General Procedure}\label{General}
We work in three spacetime dimensions, with metric tensor $\eta_{\mu\nu}= \rm{diag}(1,-1,-1)$. Consider a $ \mathcal{G}_{\lilN}\equiv\mathrm{U(1)\times...\times U(1)}$ gauge theory, where $N$ is the number of \emag factors appearing in the product that defines \lie. Let $\Phi$ be a $\mathbb{C}^{\lilN}$-valued function of spacetime coordinates, which we represent as $\Phi=(\varphi_{1},...,\varphi_{\scriptscriptstyle{\lilN}})$, where each $\vf{a}$ is a complex scalar field transforming under one \emag subgroup and coupled to a connection $ie_{a}\An{a}$. We stress that latin subscripts such as $a$ are being used simply as a way to label the different \emag subgroups of \lie, and at no point in this work summation is implied by their repetition. Throughout the paper, this index will be enclosed in curly brackets when ambiguity with Lorentz indices is not resolved by the context.

The gauge connection allows for the introduction of a covariant derivative, whose action on $\Phi$ is given by the $N$-tuple $	D_{\mu}\Phi=(D_{\mu}\varphi_{1},...D_{\mu}\varphi_{\scriptscriptstyle{\lilN}}),$ where  $D_{\mu}\vf{a}=\pb{\mu}\vf{a} + i\An{a}\vf{a}$. The coupling constants $e_{a}$ have been absorbed by the gauge fields through the redefinition $e_{a}\An{a}\to\An{a}$. An analogous operation $\overline{D_{\mu}\Phi}$ is obtained by complex conjugation of each component of $D_{\mu}\Phi$. The commutator of covariant derivatives induces a field-strength tensor, defined by its action on the Higgs field, which is of the form $F_{\mu\nu}\Phi = \left(\Ffn{\mu\nu}{1}\vf{1},..,F_{\mu\nu}^{ \{ \scriptscriptstyle{\lilN} \scriptstyle \}}\vf{\scriptscriptstyle{\lilN}}\right)$, where $\Fn{a}=\pb{\mu}\Aan{a}{\nu} - \pb{\nu}\Aan{a}{\mu}$. The field-strength tensor is used to define the fields $E_k^{\{a \}}=F_{0k}^{\{a \}}$ and $B_a=F_{21}^{\{a \}}$. To avoid confusion with the macroscopic quantity historically identified with the magnetic field, $B_a$ will be called magnetic induction, or simply induction.

The next piece of the theory is a potential $V(\{|\varphi|\})$, where the notation $\{|\varphi|\}$ stands for the set $\{|\varphi_{1}|,...,|\varphi_{\scriptscriptstyle{\lilN}}|\}$. To induce symmetry breaking, we shall choose a potential with an asymmetric minimum of the form $\{|\varphi|\}=\{v_1,...,v_{\scriptscriptstyle{\lilN}}\}$, where the $v_a$ are strictly positive real numbers. Finiteness of the energy enforces that $\Phi$ is mapped into the set $(v_1e^{i\Lambda_1},...,v_{\lilN}e^{i\Lambda_{\scriptscriptstyle{\lilN}}})$ asymptotically, and that $|D_k\varphi_{a}|^2=0$ in the same limit. These features allow for a topological classification of solutions according to the maps $\Phi_{\infty}$ between spatial infinity and the vacuum manifold. Topologically, the restriction of $\Phi_{\infty}$ to each \emag subgroup is of the form $\vfi_{\infty}:S^{1}_{\infty}\mapsto S^1$, similar to the maps encountered in MH theory. Thus, we need not perform further calculations to deduce that all finite energy solutions may be classified according to a set of $N$ integers, corresponding to the $N$ flux quantization conditions
	\begin{equation}\label{flux}
		\int d^2x B_a  =2\pi n_a.
	\end{equation}  
The topological charges $n_a$ appearing in the above equations are ubiquitous in the theory, and may in principle be deduced from indirect reasoning, such as estimations of the total mass. We note that our conventions imply that the flux quanta have the same $2\pi$ value for all sectors. These conventions do not change the unit of energy as long as each $\Fn{a}$ in the Lagrangian is multiplied by a factor of $1/e_{a}$. 

The equations of motion can be obtained from the Lagrangian density
\begin{equation}\label{Lag}
\LL=\sum_{a=1}^{\lilN}\left\{-\frac{1}{4}\frac{\Fn{a}F^{\mu\nu\{a\}}}{\mu_{a}(\{|\varphi|\})} + D^{\mu}\vf{a}\overline{D_{\mu}\vf{a}}\right \}- V(\{|\varphi|\}).
\end{equation}
This Lagrangian presents a generalization of the Maxwell term from standard scalar electrodynamics, introduced by the functions $\mu_a(\{|\varphi|\})$, which act as generalized permeabilities multiplying the observable gauge fields of the model, which behave as if propagating in a different medium. The permeabilities will be taken as bounded, nonnegative functions that do not go to zero when the fields approach the vacuum.  Geometrically, the tensor $F_{\mu\nu}$ describes the curvature of the connection. In the generalized model, this curvature is locally rescaled by the permeabilities, in what may be seen as a continuous deformation that maps the standard theory into the modified one. As mentioned above, each $\Fn{a}$ in~\eqref{Lag} must be divided by the constants $e_a$. These have been absorbed into the permeabilities, and may be recovered explicitly in~\eqref{Lag} by exchanging $\mu_a$ for $(e_a)^2\mu_a$. The symmetric energy-momentum tensor of the theory can be found directly from~\eqref{Lag}:
	\begin{align}\label{Tmn}
	T_{\alpha\beta}= &\sum_{a=1}^{\lilN}\left(-\frac{F_{\alpha\gamma}^{\{a\}} F_{\beta}^{\{a\} \gamma}}{\mu_a} + \overline{D_{\alpha}\vf{a}}D_{\beta}\vf{a} + D_{\alpha}\vf{a}\overline{D_{\beta}\vf{a}}\right) \nonumber\\
	&- \eta_{\alpha\beta} \LL.
	\end{align}
	The  field equations derived from~\eqref{Lag} are
\begin{subequations}\label{eqs}
\begin{align}
	D^{\mu}D_{\mu}\vf{a} -\frac{1}{4}\sum_{b=1}^{\lilN}\pd{\mu_b}{\overline{\vf{a}}}\frac{\Fn{b}\Fun{b}}{\mu_b^2}+ \pd{V}{\overline{\vf{a}}}= 0, \label{phieq}& \\	
	\partial^{\mu}\left(\frac{\Fn{a}}{\mu_a}\right)=J_{\nu}^{\{a\}}&, \label{Feq}
\end{align}
\end{subequations}
where $J_{\nu}^{\{a\}}\equiv i(\overline{\vf{a}} D_{\mu}\vf{a}-\vf{a}\overline{D_{\nu}\vf{a}})$. A finite energy solution of the field equations is given by $N$ sets of the form $\left(\vf{a}, \An{a}\right)$, each of which can be thought of as a collection of (anti)vortices carrying $n_a$ quanta of flux from the respective sector.

To completely specify the model, we must choose a potential, whose minimum value may always be taken as zero. The constants $v_a$ for which this minimum is achieved play no significant role in our discussion, and shall henceforth be taken as unity. As stated in the previous section, a wide range of applications could be conceived for solutions of the equations of motion~\eqref{eqs}, and this naturally must be taken into account when specifying the potential. To motivate our results, let us focus on the concrete example of a generalization from Maxwell-Higgs (MH) theory that includes hidden sectors. Those interested in applications that demand a potential with different properties may adjust the argument to find the form that best suits their needs. 

The success of the GL theory of superconductivity suggests a investigation of potentials that are reduced to the quartic GL potential for one of the $\vf{a}$, identified with the sector of ordinary electromagnetism, in the limit when all the other scalar fields tend to their vacuum values. This makes the new models consistent with known observations, since experiments performed in a region that is lacking particles of the hidden sectors will recover the results of GL theory. For definiteness, let the $N$-th sector be that of electromagnetism. The simplest way to achieve the desired behavior is by taking a potential of the form $V=V_{\lilN} + \widetilde{V}$, where $\widetilde{V}$ is left unspecified for the moment and
	\begin{equation}\label{10}
		V_{\lilN}=f_{\lambda}\frac{e_{\lilN}^2}{2}\left[(1-|\vf{\lilN}|)\right]^2,
	\end{equation}
in which the $e_{\lilN}^2/2$ factor has been chosen for convenience and $f_{\lambda}=f_{\lambda}(|\vf{1}|,...,|\vf{\lilN-1}|)$ is a function that performs a smooth deformation of the GL potential. We may interpret this deformation is a consequence of the presence of defects from sectors other than $N$. To avoid unnecessarily cumbersome equations, let us introduce the notation $\Vca{a}\equiv (e^{i\Lambda_1},...,\vf{a},...,e^{i\Lambda_{\lilN}})$, which is simply a convenient way of representing a situation where every field, with the possible exception of $\vf{a}$, has reached its vacuum value. The requirement that a GL potential is attained in the aforementioned limit can be expressed as
	\begin{equation}\label{limF}
		\lim_{\Phi\to \mathcal{V}_{\lilN}}f_{\lambda}(|\vf{1}|,...,|\vf{\lilN-1}|)=\lambda,
	\end{equation}
where $\lambda$ is the same parameter that appears in the MH model. The physical interpretation lies on the observation that $f_{\lambda}$ plays the same role in~\eqref{10} that $\lambda$ did in the GL theory, so the former is a natural generalization of the latter. It may seem strange that this role is now played by a function, but this is actually to be expected: $\lambda$ is related to many fundamental parameters in the theory of superconductivity~\cite{Kittel}, all of which are related to the response of a given material to the presence of an electromagnetic field. If $\mu_{\lilN}$ is to be seen as a magnetic permeability, than it must change the way charge carriers in the material react to a magnetic field, meaning that $\lambda$ must be changed as well. Also, $f_{\lambda}$ should change as a function of the same variables as those of $\mu_{\lilN}$, which is why the possibility that $f_{\lambda}$ could depend on quantities not belonging to $\{|\vf{a}|\}$ was not considered. Essentially,~\eqref{limF} corresponds to the requirement that $V_{\lilN}$ be identified with the standard quartic potential when no non-vacuum fields from \textit{different} sectors exist in a given region.

To advance, we must make some assumptions about the hidden sectors. A natural way to proceed is to extend~\eqref{10} to other sectors, leading to the potential 
	\begin{equation}\label{pot}
		V=\sum_{a=1}^{\lilN}V_a\equiv\sum_{a=1}^{\lilN}f_{\lambda_a}\frac{(e_a)^2}{2}\left[(1-|\vf{a}|)\right]^2,
	\end{equation}
where each $f_{\lambda_a}=f_{\lambda_a}(|\vf{1}|,...,|\vf{\lilN-1}|)$ is a function defined in analogy to $f_{\lambda}$ (which we have renamed $f_{\lambda_{\lilN}}$ in~\eqref{pot}). A straightforward adaptation of~\eqref{limF} holds when $\Phi\to\mathcal{V}_a$, with $\lambda$ replaced by $\lambda_a$. By defining the potential~\eqref{pot} we are assuming that the physics of the hidden sectors is sufficiently analogous to that of the visible one for a model similar to the MH theory to be sensible.

Eq.~\eqref{pot} leads to a Lagrangian of the form $\LL=\LL_1 + ... + \LL_{\lilN}$. Although the additive factors $\LL_a$ are coupled and therefore not independent of each other, they are, in a sense, representative of the respective sectors after which they have been labeled. Indeed, each of them gives rise to a MH-like action when the vortices of different sectors are well separated. To make this statement more precise, recall that the vortex positions may be identified with the zeros of the scalar fields, which we assume are isolated. For each $a$, let $\gamma_a$ be a loop enclosing all the zeros of $\vf{a}$. If these loops are at a sufficiently great distance from each other, then, as we know, $V_a$ is just the quartic potential of MH theory. The permeabilities $\mu_a$ must also become independent of $|\vf{b}|$ for all $b\neq a$ in this limit, since $\left(\vf{b}, \An{b}\right)$ is a defect located in $\gamma_b$ and which, as such, must have reached the vacuum at the interior of $\gamma_a$. The restriction of $\LL_a$ to $\gamma_a$ is therefore a function of $\vf{a}$ and $\An{a}$ only. In particular, it becomes the MH Lagrangian if $\mu_a$ is independent of $\vf{a}$. If these loops, along with the zeros enclosed within, are continuously moved close to each other, the $\LL_a$ are smoothly deformed into the modified models this paper concerns itself with. Since $\LL_a$ depends only on functions of the $a$-th sector in the aforementioned limit, it is sensible to identify this Lagrangian with $\vf{a}$ and $\An{a}$.

Next, we must specify $f_{\lambda_a}$ and $\mu_a$. There is freedom in these choices, which may be used to allow for a Bogomol'nyi bound. The existence of this bound for some combination of parameters provides a powerful tool of analysis, with use of which many results from vortex theory have been proved, and which was even instrumental in the analysis of vortex stability \textit{outside} critical coupling~\cite{Gustafson}. Moreover, the BPS limit is an invaluable asset in investigations related to low energy dynamics of vortices through the adiabatic approximation. This approach stems from the realization, due to Manton~\cite{geodesic} in the context of monopole scattering, that the low speed dynamics of defects can be approximated by geodesic motion on the Moduli space generated by solutions of the BPS equations. This approximation has been successfully applied to vortices and other topological defects~\cite{Ruback,geodesicII,Speight,ADI,Impurities,Palvelev,Fuertes}. In~\cite{Stuart}, Stuart proved the adiabatic approximation rigorously for the MH theory, and showed that it also holds when $\lambda= 1+\epsilon$, where $\epsilon$ is a small constant quantifying deviation from critical coupling. 

As in the MH theory, each additive term in~\eqref{pot} is in fact a family of potentials, parametrized by $\lambda_a$. In the MH model, there is an attractive force between the zeros of $\vf{a}$ when $\lambda_a < 1$, and a repulsive one when $\lambda_a>1$. The physical meaning of the BPS limit lies in the vanishing of this force in the limit where $\lambda_a\to 1$. As the MH Lagrangian is deformed into $\LL_a$, new forces emerge between vortices of \emph{different} sectors, but we suppose that a generalized permeability does not change the direction of forces occurring between vortices of the \emph{same} sector, suggesting that the boundary between attractive and repulsive forces should still be defined by $\lambda_a=1$. As will be verified shortly, these conditions will lead to a Bogomol'nyi bound when
\begin{equation}\label{rel}
(e_a)^2f_{\lambda_a}=\lambda_a\mu_a, 
\end{equation} 
where the $(e_a)^2$ factor appears as a consequence of our earlier redefinition, which lead to $(e_a)^2$ being absorbed into the permeabilities. Together with~\eqref{limF}, this equation implies $\mu_a\to (e_a)^2$ at infinity, so that at great distances from the vortices, all observable fields obey Maxwell's equations for a medium of constant permeability. Thus, the effect of vacuum solutions in the generalized permeability is at most that of a constant background. For notational simplicity, we shall henceforth take $e_1=...=e_{\lilN}=1$ in our discussion. These constants may be recovered through the transformations $\mu_a\mapsto (e_a)^{2}\mu_a$. 

An interesting realization, and one not \textit{a priori} obvious, is that our argument has also lead us, in a natural way, to linear equations relating $f_{\lambda_a}$ and $\mu_a$. This is the simplest possible nontrivial relationship between these functions. The physical reasoning used to motivate~\eqref{rel} implies that $f_{\lambda_a}$ is independent of $\lambda_a$ at least when non-vacuum fields are absent. This condition may be relayed in exchange for a bit more generality,  as allowing for $|\vf{a}|$ dependence on $f_{\lambda_a}$ (and therefore $\mu_{a}$) invalidates none of our main results. On the one hand, a generalization of this kind would change (or possibly eliminate) the relationship between our generalized model and a MH-like Lagrangian from the $a$-th sector. On the other hand, the extra freedom gained by dropping this assumption allows for some interesting choices, among which is the $|\varphi|^6$ potential, that amounts to  $f_{\lambda_a}=\lambda_a[\alpha^2 + (|\vf{a}|)^2-1]$. This potential is frequently used in theories with a Chern-Simons term~\cite{Jackiw}. In fact, a model with $df_{\lambda_a}/d|\vf{a}|\neq0$ for some $a$ will be presented in Sec.~\ref{models}, although we shall still require that~\eqref{limF}, in the form it is stated, holds for at least one sector, identified with standard electromagnetism. As before, we shall conventionally label it by $N$. Whenever the association with the GL or MH theory is relevant to the interpretation of a result from this paper, we will state it explicitly. 
 
Let $\lambda_a=1 + \epsilon_a$, where $\epsilon_a$ is a constant for each $a$. If  $\mu_a$ has zeros, one must assume that there is at least one nontrivial configuration such that the limit of $\Fn{a}F^{\mu\nu \left\{a\right\}}/\mu_a$ exists at every point in the plane.  When~\eqref{pot} and~\eqref{rel} hold, the energy becomes
	\begin{equation}\label{15}
		\begin{split}
			{E} &=\sum_{a=1}^{\lilN}\int d^2x\left\{  \left(|D_0\vf{a}|^2 + \frac{1}{2\mu_a}E_k^{\{a \}}E_k^{\{a \}}\right) \right.\\
			&\bigg.+ |D_1\vf{a} \pm iD_2\vf{a}|^2 + \frac{\epsilon_a}{2}\mu_a (1 - |\vf{a}|^2)  \\
			&\bigg.+ \frac{1}{2\mu_a}\left[\Bn{a}\mp \mu_a(1- |\vf{a}|^2)\right] ^2  \pm B_a\bigg\}. 
		\end{split}
	\end{equation}

If we let $\epsilon_a\to 0$ for all $a$ and use~\eqref{flux}, we are led to the bound
	\begin{equation}\label{Energy}
		{E} \geq 2\pi\sum_{a=1}^{\lilN}|n_a| ,
	\end{equation}
 which is saturated when every quadratic term in~\eqref{15} is zero. As is well known, we may use our gauge freedom to impose $A_0^{\{a\}}=0$ as long this is compatible with Gauss's law~\cite{coleman}. With this gauge choice, the conditions $D_{0}\vf{a}=E_k^{\{a \}}=0$ imply $\pb{0}\vf{a}=0$. The energy is minimized by solutions of the first order equations 
	\begin{subequations}\label{BPS}
		\begin{align}\label{BPSa}
		&B_a=\pm\mu_a(1-|\vf{a}|^2), \\
		&(D_1 \pm iD_2)\vf{a}=0.\label{BPSb}
		\end{align}
	\end{subequations}

The upper (lower) signs in these equations give vortex (antivortex) solutions. They are related by mapping $\vf{a}\mapsto\overline{\vf{a}}$ and $A_k^{\{a\}}\mapsto-A_k^{\{a\}}$. For this reason, only the upper signs will be kept in our discussion.  It may be directly verified from~\eqref{Tmn} that BPS solutions satisfy the condition $T_{ij}=0$. Hence, one may integrate the continuity equation $\pb{0}T_{0j} + \pb{i}T_{ij}=0$ over a volume to verify the absence of forces for static solutions of~\eqref{BPS}. An interesting geometrical interpretation is possible when all the $\mu_{a}$ are independent of $|\vf{\lilN}|$, in which case the first $N-1$ sets of equations in~\eqref{BPS} may be solved independently. In that case, the generalization introduced by $\mu_a$ amounts, in the BPS limit, to multiplication of~\eqref{BPSa} by a function $\mu_{\lilN}(x,y)$ of the coordinates. Equations of this form are encountered in investigations of vortices in curved Riemannian manifolds~\cite{Bradlow, Edelstein, Prada, Romao}, where the function $\mu_{\lilN}(x,y)$ plays the role of a metric conformal factor.

Existence and uniqueness of solutions has been proved for a variety of cases when $\mu_a$ is a function of $|\vf{a}|$ and the coordinates only, see for example~\cite{Taubes, Lohe, CS,ChenI, ChenII, Wang}, and also~\cite{hyperbolicI,hyperbolicII} for some exact solutions. In the general case investigated here, we shall be content to assume existence, and present some explicit solutions in Sec.~\ref{models}. 

The scalar fields of BPS solutions solve 
	\begin{equation}\label{DkDk}
		D_kD_k\vf{a}=\mu_a(1-|\vf{a}|^2)\vf{a},
	\end{equation} 
which, when $\mu_a$ is independent of $\vf{a}$, has locally the form of the static MH equation. Suppose that, inside a small disk centered at a point $P$, $\mu_{a}$ changes slowly, so that it may be approximated to any desired accuracy by $\widetilde{\lambda}\equiv \mu_{a}\big|_{P}$, provided the radius of the disk is small enough. Thus, the change in $\vf{a}$ is governed by a GL equation with parameter $\widetilde{\lambda}$.  $B_a$ solved, in the same neighborhood, the appropriate Maxwell equations for a medium of constant magnetic permeability $\mu_{a}\big|_{P}$. If $\widetilde{\lambda} < 1$, the system behaves as a type-I material, while $\widetilde{\lambda} > 1$ implies a type-II material. Thus, the magnetic properties will change throughout the plane, as different regions of the same system may behave as either type of material. When, in particular, there exists a disk inside which $\mu_a \approx 0$, $B_a$ will be expelled from that neighborhood, as would happen in a superconductor.

Equations~\eqref{BPSa} also show that zeros of $\mu_a$ generate zeros of the magnetic field.  Assuming a smooth solution to the first-order equations exists, one may use~\eqref{BPSb} and the Poincaré Lemma argument from~\cite{TaubesII} to prove that the fields are holomorphic functions of $z=x + iy$, with a finite number of zeros. This implies that a neighborhood  of any zero of $\vf{a}$ may be mapped to 		
	\begin{equation}\label{zeros}
		\vf{a}=\phi_{a}(z)(z-z_l)^{m_l},
	\end{equation} 
where $\phi_a(z_l)\neq 0$ and $m_l$ is a positive integer. Had we taken the lower signs in~\eqref{BPS}, we would have instead found a pole of order $m_l$. In the models presented in Sec.~\ref{models}, we shall specialize in permeabilities that possess either a minimum $\mu_{min}<<1 $ or a maximum $\mu_{max}>>1$ at zeros of some $\vf{a}$. With this condition, the model is strongly deformed (in relation to the \qt{standard} $\mu_a=1$ case) at the positions of the vortices, which may be interpreted as causing the deformation. Let us again identify the $N$-th sector with electromagnetism and, for simplicity, suppose that, at least in a region of the plane, $\mu_{\lilN}$ may be treated as a function of a single field $\varphi_1$, with a minimum for $\varphi_1=0$.  These zeros generate neighborhoods where the system behaves as a type-I superconductor, as the magnetic induction is expelled. The larger the order $m_l$ of this zero, the larger the disk in which $(z-z_1)^{m_1}\approx 0$ holds to a given approximation. This exemplifies how vortex density at a given point influences the rate at which the generalized permeabilities grow out of a minimum. If $\mu_{\lilN}$ has instead a pronounced maximum at $\varphi_1=0$, the inverse behavior is found, as this neighborhood allows more magnetic flux in proportion to $\mu_{max}$.

When~\eqref{BPS} are solved, the energy is at a global minimum, and it should be noted that this happens when the functionals $E_a$ defined by spatial integration of $\rho_a=-\LL_a$ are each minimized. As expected, the set ${n_1,...,n_{\lilN}}$ completely specifies the energy of a BPS solution. This observation may be used to predict the flux produced by hidden sectors. For example, in the $N=2$ case we could have one \emag symmetry for electromagnetism and one for the analogue interaction of a hidden sector. One could find the energy $E_2$ of the visible sector by measurement of the magnetic flux. With this information, any reasonable estimation of the total energy would allow an experimentalist to indirectly determine both the energy $E_1$ and flux relative to this hidden sector. 

For minima of the energy, this reasoning may work even when the bound~\eqref{Energy} is not exactly saturated: since a small change in $\lambda_a$ produces a small variation of the energy functional, we expect that $\lambda_a=1$ is a good approximation to $\lambda_a\simeq1$. In that case, solutions of~\eqref{BPS} may be used to study vortices near critical coupling, thus allowing for an estimation of the topological sector even outside the Bogomol'nyi limit. Indeed, if $|\epsilon_a| << 1$, then the term proportional to $\epsilon_a$ in~\eqref{Energy} will amount to a very small fraction of the potential energy, so it appears reasonable that the Bogomol'nyi bound may be used as an estimation to the energy in this approximation. For the standard case, a thorough analysis of vortex dynamics near critical coupling was conducted in~\cite{Stuart}, showing that deviation from the Bogomol'nyi regime when $\lambda=1+\epsilon$ results in an interaction potential that is small when $\epsilon$ is. An attempt to replicate this rather involved demonstration will not be made at this time, but we see no impediment to generalization of his results to the present case, since the author was able to generalize his argument even to the (mathematically very distinct) case of magnetic monopoles~\cite{StuartII}. Although new forces may appear in our theory because of the different species of vortices, these would also be expected to be small near the Bogomol'nyi limit. 

\section{Rotationally symmetric vortices}\label{sym}

Now we turn our attention to rotationally symmetric solutions, as has been the standard procedure since Abrikosov~\cite{abrikosov}. We will investigate time-independent solutions with an isolated zero at the origin, and take $A_0=F_{0k}=0$. Up to gauge transformations, the most general~\cite{manton} solutions of this kind have the form
\begin{subequations}\label{Ansatz}
	\noindent\begin{minipage}{.5\linewidth}
		\begin{equation}\label{g}
		\vf{b}=g_b e^{in_b\theta},
		\end{equation}
		\vspace{0.5pt}
	\end{minipage}%
	\begin{minipage}{.5\linewidth}
		\begin{equation}\label{a}
		\vec{A}^{\{b\}}=\frac{n_{b}-a_{b}(r)}{r}\hat{\theta},
		\end{equation}
		\vspace{0.5pt}
	\end{minipage}
\end{subequations}
\noindent where, for simplicity, we have assumed solutions symmetric with respect to the same point. Fields of this form lead to a problem with fewer and simpler differential equations. But in this paper there is another advantage to their use: these solutions have all their zeros at the origin, and are thus the perfect setting for our proposed goal of studying the role of the winding numbers in the generalized theories, as these invariants appear manifestly in~\eqref{Ansatz}. However, this ansatz must be complemented by the boundary conditions
\begin{subequations}\label{conds}
	\noindent\begin{minipage}{.5\linewidth}
		\begin{align}\label{gbond}
			&g_b(0) =0, \\ 	
			&g_b(\infty)=1 ,	
		\end{align}
		\vspace{0.5pt}
	\end{minipage}%
	\begin{minipage}{.5\linewidth}
		\begin{align}\label{abond}
		&a_b(0) =n_b,  \\
		&a_b(\infty)=0.	
		\end{align}
		\vspace{0.5pt}
	\end{minipage}
\end{subequations}
These solutions have a zero with multiplicity $n_b$, which may be interpreted as the number of vortices from the $b$ sector located at the origin. We shall continue to assume $n_b>0$, since the antivortex solutions are analogous.  For fields of the form~\eqref{Ansatz}, the equations of motion become
 \begin{subequations}\label{SO}
	\begin{align}
		\begin{split}
			&\frac{1}{r}\left(r g_b'\right)'=\frac{g_ba_b^2}{r^2} + \mu_b\lambda_b(1-g_b^2)g_b \\
			&\ -\frac{1}{4}\sum_{c=1}^{\lilN}\pd{\mu_c}{g_b}\left[\left(\frac{a_c'}{\mu_cr}\right)^2  - \lambda_c\left(1-g_c^2\right)^2\right],\label{SO1}
		\end{split}
		\\[1ex]
		&\left(\frac{a_b'}{\mu_br}\right)' = \frac{2a_b g_b^2}{r}, \label{SO2}
	\end{align}
 \end{subequations}
where the prime denotes differentiation with respect to the radial coordinate. Solutions of these equations have a total energy density $\rho=\sum_{b=1}^{\lilN}\rho_b$, with
 \begin{equation}\label{rho}
		\rho_b= \frac{(B_{b})^2}{2\mu_b} +  \mu_b\frac{\lambda_b}{2}(1- g_b^2)^2 +g_b'^2 +\left(\frac{g_ba_b}{r}\right)^2,
 \end{equation}
 where $B_b=-a_b'/r$ and $f_{\lambda_b}=\mu_b\lambda_b$. In the situation in which $\mu_b$ is independent of $g_b$, the above expression has the form of a MH energy density modified by the transformations $B_b\to B_b/\sqrt{\mu_b}$, $\lambda_b\to \mu_b\lambda_b$. This generalization has important consequences for stability, as the partial derivatives of $\mu_b$ appear in the Hessian of the energy, potentially changing the signs of its eigenvalues. In particular, a stable circular vortex with $\lambda_{b} > 1$ and $n_b>1$ may be possible when different species of vortices are superimposed. As is well known, this does not happen in the GL theory~\cite{Gustafson}, to which $E_b$ corresponds at large distances from other sectors. We owe this new theoretical possibility to the couplings introduced by the permeabilities which, outside of the Bogomol'nyi limit, cause new forces to appear between zeros of different sectors. If the net force is attractive for some combination of parameters, the symmetric solution is expected to be an energy minimizer.

 When the potential allows for a Bogomol'nyi bound, the upper sign equations in~\eqref{BPS} become
	\begin{subequations}\label{FO}
		\noindent\begin{minipage}{.5\linewidth}
			\begin{equation}\label{bps1}
		  		g_b' =\frac{a_bg_b}{r}, 
			\end{equation}
			\vspace{0.5pt}
		\end{minipage}%
		\begin{minipage}{.5\linewidth}
			\begin{equation}\label{bps2}
			-\frac{a_b'}{r}=\mu_b(1-g_b^2).
			\end{equation}
			\vspace{0.5pt}
		\end{minipage}
	\end{subequations}
The above equations imply $g_b\propto r^{n_b}$ near the origin, which is the form taken by~\eqref{zeros} when all zeros are located at the origin. The way symmetric solutions behave near the center of the defect is important to our analysis, so let us investigate it under as general assumptions as possible. We assume that the equations of motion are well defined in a neighborhood containing the origin, meaning in particular that functions such as $a_b'/(r\mu_b)$ and $g_b'$ have a well defined limit at $r=0$. In fact, we demand that $g_b$ has a bounded derivative at least in a neighborhood of the origin. Formally, we assume that for an arbitrarily small $\delta>0$, there exists a positive number $M_{b}$ such that, for every $r$ in $U_{\delta}\equiv(0,\delta)$, $|g'(r)|\leq M_{b}$. This condition is in fact satisfied by any real function with a continuous derivative in the interval~\cite{Zygmund}. By the mean value theorem, it also implies
\begin{equation}\label{ineq}
		g_b(r)\leq M_{b}r,
\end{equation}
for all $r\in U_{\delta}$. For $r\neq 0$, we may take the square of inequality~\eqref{ineq}, divide by $r$ and integrate the result over $(0,r)\subset U_{\delta}$ to find
	\begin{equation}\label{intineq}
			\int_{0}^{r}\frac{g_b^2}{r}dr \leq	\frac{M_{b}^2}{2}\delta^2,
	\end{equation}
where we used the fact that both $g_b$ and $M_b$ are nonnegative functions, so that the inequality is preserved. By~\eqref{abond} and continuity, one may write $a_b=n_b-\alpha_b$, with $\alpha_b$ small in $U_{\delta}$. Integrating~\eqref{SO2} over that region and using~\eqref{intineq}, we find
	\begin{equation}\label{hb}
		h_b(r) - h_b(0) \simeq 2n_b\int_{0}^{r}\frac{g_b^2}{r}dr\leq n_bM_{b}^2\delta^2,
	\end{equation}
where $h_b\equiv -B_b/\mu_b$, and $h_b(0)$ is the limit of this function as $r\to 0$, which exists by hypothesis. As $\delta$ becomes arbitrarily small, the right hand side of this inequality vanishes. We shall return to~\eqref{hb} soon, but for now note that, since $h_b(0)$ is a real number, $h_b(r)$ must be bounded in $U_{\delta}$. 

Let us now turn our attention to the remaining equation in~\eqref{SO}. As in the MH theory, the leading order contribution from the first expression in~\eqref{SO1} is $g_bn_b^2/r^2$. The second term in that equation gives a leading contribution $\mu_n\lambda_bg_b$, which cannot dominate over the first one unless $\mu_c$ diverges faster then $r^{-2}$ in $U_{\delta}$. This cannot happen for a bounded $\mu_c$. We have assumed the zeros of $\vf{b}$ are isolated, thus, for $r\neq 0$, we may multiply this equation by $r^2/g_{b}$ to find
	\begin{equation}\label{27}
		\begin{split}
	\frac{r}{g_b}\left(rg_b'\right)'= n_b^2 -\sum_{c=1}^{\lilN}\frac{r^2}{g_b}\pd{\mu_c}{g_b}\left\{\frac{[h_c(0)]^2 -\lambda_c}{4}\right\},
	\end{split}
	\end{equation}
where only the smallest order coefficients were kept. Since $h_c(0)$ (and therefore its square) as well as $\lambda_c$ are real numbers, the leading order term will ultimately depend on how the functions
	\begin{equation}
		\chi_{c}^b\equiv \frac{r^2}{g_b}\pd{\mu_c}{g_b}
	\end{equation}
\noindent behave when $r\to 0$. Since $n_b^2$ is a zeroth-order expression, the \qt{BPS-like} leading order behavior $g_{b}\propto r^{n_b}$ is recovered from Eq.~\eqref{27} if, for some $\delta>0$, $\chi_{c}^b=0$ holds for all $r\in U_{\delta}$ to at least order one. A sufficient condition for this to happen is that, for every $c$, there exists a nowhere divergent function $\zeta_{bc}(\{ |\varphi|\})$ such that
	\begin{equation}\label{righthandlimit}
		\lim_{|\vf{b}|\to 0^+}\frac{1}{|\varphi_b|}\pd{\mu_c}{|\vf{b}|}=\zeta_{bc}(\{ |\varphi|\}).
	\end{equation}
 This condition is, in particular, satisfied for any twice differentiable permeability such that $\partial\mu_c/\partial|\varphi_b|$ vanishes as $|\vf{b}|$ approaches zero. If instead some of the $\chi_{c}^b$ diverge or tend to a nonzero real number, the behavior of $g_b$ may also depend on winding numbers other than $n_b$, as the last term in~\eqref{27} will need to be taken into account. For permeabilities that are multivariate polynomials in the fields, or that may be treated as such in a neighborhood of $\{\varphi \}=\{0,...,0 \}\equiv \{0\}$, Eq.~\eqref{righthandlimit} is always satisfied if a linear term in $|\vf{b}|$ is absent.

As an example, consider a $\rm{U(1)}\times\rm{U(1)}$ model defined by generalized permeabilities $\mu_1=\mu_1(|\vf{1}|)$ and $\mu_2=\mu_2(|\vf{1}|, |\vf{2}|)$, whose derivatives with respect to $g_1$ satisfy~\eqref{righthandlimit}. In that case, Eq.~\eqref{27} implies $g_1=C_1r^{n_1}$ near the origin. Moreover, let $\mu_2$ be a polynomial function in $|\vf{1}|$ and $|\vf{2}|$. Under these assumptions, any contribution that might lead to $\chi^{2}_{2}\neq 0$ at the origin must come from terms linear in $|\vf{2}|$. Therefore, it suffices to consider permeabilities that may be approximated, at least in a neighborhood of $\{0\}$, by a function of the form $\mu_2=|\vf{1}|^{\alpha}|\vf{2}|$, where $\alpha$ is a nonnegative integer different from one, thus leading to $\chi^{2}_{2}\propto r^{\alpha n_1 + 2}/g_2$ in that neighborhood. For $n_2\neq \alpha n_1 + 2$, substitution of this result in~\eqref{27} with $g_2(0)=0$ gives
	\begin{equation}\label{30}
		g_2=C_2r^{n_2} -\frac{1}{4}\frac{(C_1)^{\alpha}(h_1^2(0)- \lambda_1)}{(\alpha n_1 + 2)^2 - n_2^2}r^{\alpha n_1 + 2},
	\end{equation} 
where $C_1$ is not a new integration constant, but the one appearing in $g_1$. For $n_2< \alpha n_1 + 2$ we find, to leading order, $g_2=C_2r^{n_2}$. Note that $\alpha$ is nonnegative, so this inequality is always satisfied for $n_2=1$. If, on the other hand, $n_2> \alpha n_1 + 2$, then $g_2\propto r^{\alpha n_1 + 2}$. When $n_2$ and $\alpha n_1 + 2$ are equal, the second term in Eq.~\eqref{30} must be replaced by a contribution proportional to $\ln(r)r^{n_2}$. We see that the behavior of $g_2$ near the origin depends on the relationship between $n_1$ and $n_2$. Interestingly, increasing $n_2$ past the value $\alpha n_1 + 2$ does not change the power with which $g_2$ grows out of its zero. This example illustrates the way winding numbers from different sectors may ultimately define the leading order behavior of non-BPS symmetric vortices, as the $g_b\propto r^{n_b}$ may become impossible by virtue of the partial derivatives in~\eqref{SO1}. While interesting, it seems unlikely that a solution with such alternative behavior would be a minimum of the energy, at least when all $\lambda_c\simeq 1$, since one would expect a minimizer to mirror the behavior of a BPS solution when critical coupling is approached. Thus, let us assume~\eqref{righthandlimit} holds for the remainder of this section.

We note that the analysis developed in this section may be generalized to include models that fail to obey~\eqref{rel}, in which case the partial derivatives of $f_{\lambda_c}$ must be subject to the same investigation we have conducted for those of $\mu_{c}$. One important example of this kind arises if one wishes to keep unchanged the quartic potential, which is well understood and often used in quantum field theories. In this case, one would introduce a generalization only in the permeabilities, which would not be related to $V(\{|\varphi|\})$. The leading order contribution would still depend only on $\chi_c^b$, as this modification would amount to setting $f_{\lambda_c}=1$ and removing the last additive term in the right hand side of Eq.~\eqref{SO1}.

In contrast to the Nielsen-Olesen (NO) solution, $B_b$ need not have a maximum at the origin. For sufficiently small $\delta$, the right hand side of~\eqref{hb} can be neglected, so that $-B_b/\mu_b\simeq h_b(0)$ inside $U_{\delta}$. By hypothesis, $h_b(0)$ is a real number, so $B_b\propto \mu_b$ in this neighborhood. Since any symmetric vortex has a zero at the origin, $\mu_b$ can be approximated by its value at $ \{0\}$, to an order depending on the winding numbers. To make this statement more quantitative, assume $\mu_b$ is continuous on a closed domain with boundary $\{0\}$, so that it may be uniformly approximated by a polynomial in $g_1,...,g_{\lilN}$~\cite{Stone}. Together with  $g_c\propto r^{c}$, these considerations make it possible to write ($\mu_b - \mu_b(\{0\})) \propto r^{\alpha_1n_1 +...+\alpha_{\lilN}n_{\lilN}}$ to leading order, where $\alpha_{i}$ are nonnegative integers and use was made of the fact that a polynomial of this kind is a linear combination of monomials, i.e., products involving some of the $g_c$. 

In particular, if $\mu_b(\{0\})=0$, we find $B_b=0$ at the origin. A symmetric NO solution is composed of a core of normal state, at the center of which the magnetic induction is greatest, surrounded by superconducting material. Here, we see that the opposite behavior may be found, as the generalized permeability may force $B_b$ out of a disk centered at the origin, as in a core made of a type-I superconducting material. The size of that disk depends on the winding numbers, which affect the growth of $\mu_b$. As $r$ increases, $\mu_{b}$ will eventually cease to be negligible, and the fields will grow out of that core. In order to fulfill the boundary conditions, $B_b$ must therefore have at least one maximum. Conversely, one may have a maximum $\mu_b(\{0\})>> 1$, which will result in $B_b$ peaked at the origin, but with a peak larger than that of the NO vortex. In that case, $B_b$ will be trapped at its maximum over a region around the origin.

The two first contributions in~\eqref{rho} are of order $\mu_b$ near the origin, so that the same considerations of the previous paragraphs apply. The last two however, are proportional to $r^{n_b}$, and thus their near-origin behavior at the qualitative level is determined by the winding number corresponding to its own sector, precisely in the manner of the NO vortex. When $|n_b|=1$, the origin cannot be a minimum of $\rho_b$, for the last two terms are nonzero constants to leading order. If however, $|n_b|\neq1$ the energy density may have a zero at this point, which does not happen in the MH theory, regardless of the topological charge~\cite{manton}. Here, this possibility may arise from a combination between the topological invariants $n_c$, $c\neq b$, which control the behavior of $\mu_b(\{0\})$ in a theory that can be reduced to GL superconductivity, and its own winding number $n_b$, which makes the last two terms go to zero when two or more identical vortices are superimposed.

As mentioned above, the stability of symmetric solutions depends on the forces between the vortices superimposed at the origin, and may in general depend on the specific choices for the permeability functions as well as the values of $\lambda_a$. If vortices are superimposed at the origin to give, at a given instant, a configuration of the form~\eqref{Ansatz}, the results of this section will be valid. If the net forces between zeros are attractive, this configuration would be expected to remain stable over time. The opposite scenario is interesting as well: consider for example a magnetic permeability $\mu_{\lilN}$ that has a zero for $\varphi_1=0$, and let $n_1>1$. A repulsive force may scatter the $n_1$ zeros of $\varphi_1$ in the plane. At low speeds near the Bogomol'nyi limit, we may use Manton's argument~\cite{geodesic} to model the ensuing evolution as geodesic motion in the Moduli space. Since~\eqref{BPS} is satisfied throughout this motion, the magnetic induction will correspondingly acquire new zeros in the vortex positions. We will leave this matter open for future investigation and shall instead solve~\eqref{FO}, which always give a minimum of the energy. These solutions may either be seen as an approximation valid for an energy minimizer with $\lambda_{a}\approx 1$ for all $a$ or as exact BPS solutions.

\section{Models}\label{models}

As will be shortly exemplified, solutions encountered in $\mathcal{G}_{\lilN-1}$ models are in a sense contained in $\mathcal{G}_{\lilN}$ theories. Indeed, BPS sectors (which are specified by the set $(n_1,...,n_{\lilN})$) possessing one of the winding numbers equal to zero give rise to solutions that are equivalent to those from $\mathcal{G}_{\lilN-1}$ models. This happens because a solution of~\eqref{BPS} with $n_a=0$ does not give rise to any observable quantities from that sector at the classical level, and does not affect the remaining equations of motion. In particular, consider a $\mu_b$ that is independent of $|\vf{b}|$ (recall that we argued in Sec.~\ref{General} that this condition must be satisfied at least by $\mu_{\lilN}$). Now recall that vacuum fields have no effect on the permeability. Thus, if $|\vf{a}|=1$ for every field $\mu_b$ is a function of, then $\mu_b=1$. Note that there is always a BPS sector in which this condition is satisfied. In that sector, both the first-order equations for the $b$-th sector and the corresponding energy density $\rho_b$ are identical to those of a critically coupled MH system. This implies that in any model satisfying this condition, there exist  BPS solutions that include NO vortices of the $b$-th sector~\footnote{Because this argument does not rely on rotational symmetry, we may in fact infer the presence of the multivortex solutions that have been proved to exist in critically coupled MH systems~\cite{Taubes}, with zeros at arbitrary points of the plane}. Such a feature is a consequence of the BPS limit, by virtue of which partial derivatives of the form $\partial\mu_c/\partial|\varphi_b|$, with $b\neq c$, vanish from the equations of motion. Let us now consider some $N=3$ models that illustrate the properties described above.

\subsection{Quadratic permeabilities and $\varphi^4$ potential}\label{first}
In our first example, let us take the permeabilities in as simple a form as possible, allowing us to focus on the way the vortex densities affect our results. Thus, let 
	\begin{align}\label{ex1}
		\mu= (1, |\varphi_1|^2, |\varphi_2|^2)
	\end{align}
where $\mu\equiv (\mu_1, \mu_2,\mu_3)$. The first choice gives the usual BPS solution from the standard theory~\cite{Nielsen, bogo}. As explained above, this choice is quite natural, as it may correspond to a BPS sector of a higher symmetry model, where $\mu_1$ may be a function of an unspecified field, becoming unity when that field has zero winding. $\mu_2$ and $\mu_3$ have minima respectively at the zeros of $|\varphi_1|$ and $|\varphi_2|$ which, in the symmetric case, are all located at the origin. The resulting vortices expel the corresponding induction lines from the system, in stark contrast to an Abrikosov vortex, at the center of which the magnetic field is peaked. For this configuration, $\mu_2$ ($\mu_3$) only differs from zero when terms of order  $2n_1$ ($2n_2$) are taken into account. Therefore, as the winding from the previous sector increases, so does the neighborhood in which $\mu_a\simeq 0$ holds for these permeabilities. Moreover, in symmetric solutions, $|\varphi_a|$ typically increases monotonically to their vacuum values as the distance from the origin increases. This rule, to which our fields are no exception, implies that monotonic functions of $|\vf{a}|$ become monotonic functions of $r$ in the symmetric case. Thus, at $r=0$, the vortex from the third sector (which is identified with standard electromagnetism) looks as different as possible from those found in GL theory, as $\mu_3$, and therefore $f_{\lambda_3}$, are as far from unity as possible. Another feature of this model is that $\partial\mu_b/\partial|\vf{c}|=0$ for all $c\geq b$. Thus we may solve the equations of the first sector independently, and use the result to write $\mu_{2}$ as a function of $r$. As explained in Sec.~\ref{General}, the resulting BPS equations are identical to those of a MH system in a two-dimensional Riemannian manifold whose metric is defined by a rotationally symmetric conformal factor $\mu_2(r)$. The same is of course true for the third sector.

\begin{figure}[h]
		\centering
		\includegraphics[width=4.2cm]{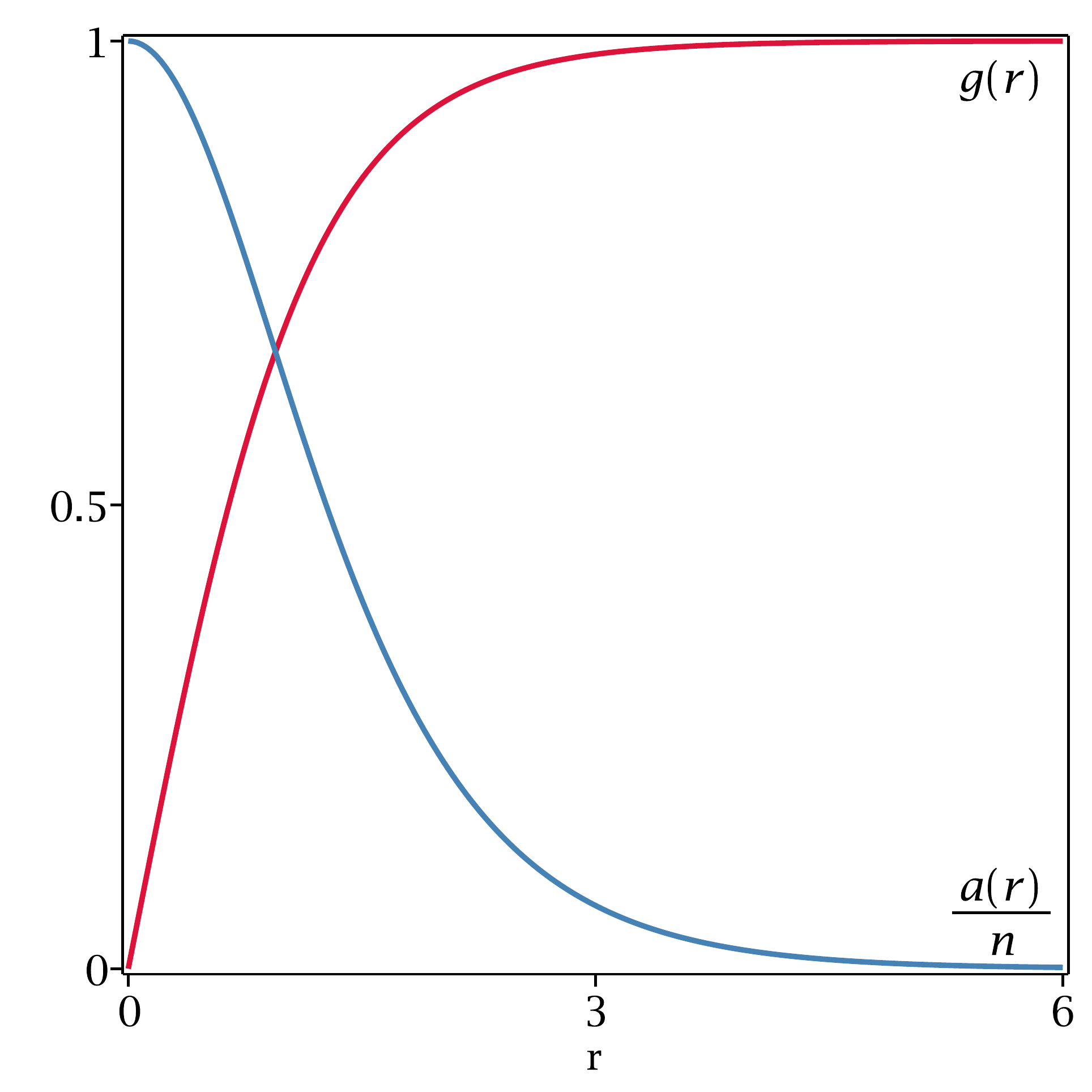}
		\includegraphics[width=4.2cm]{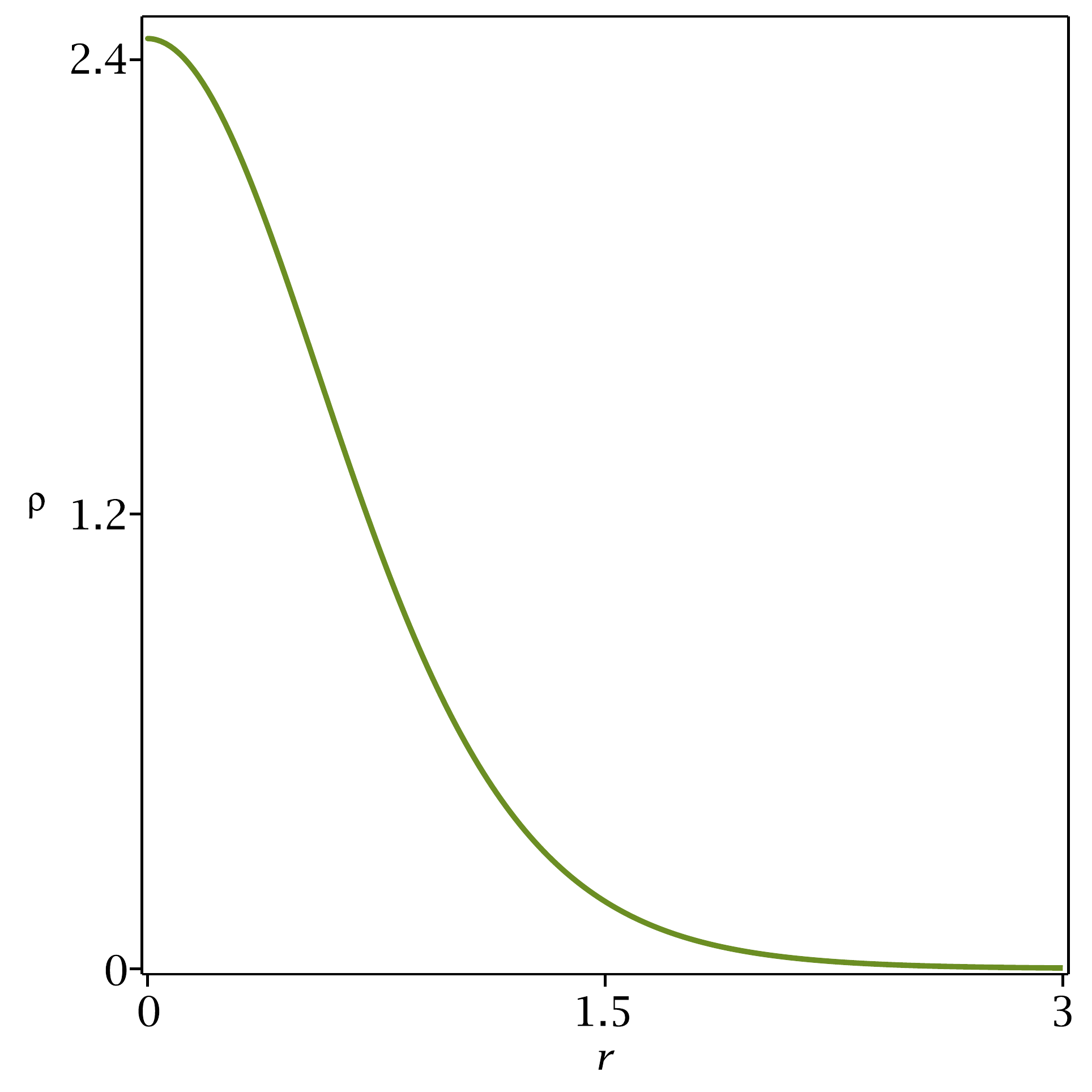}
		\includegraphics[width=4.2cm]{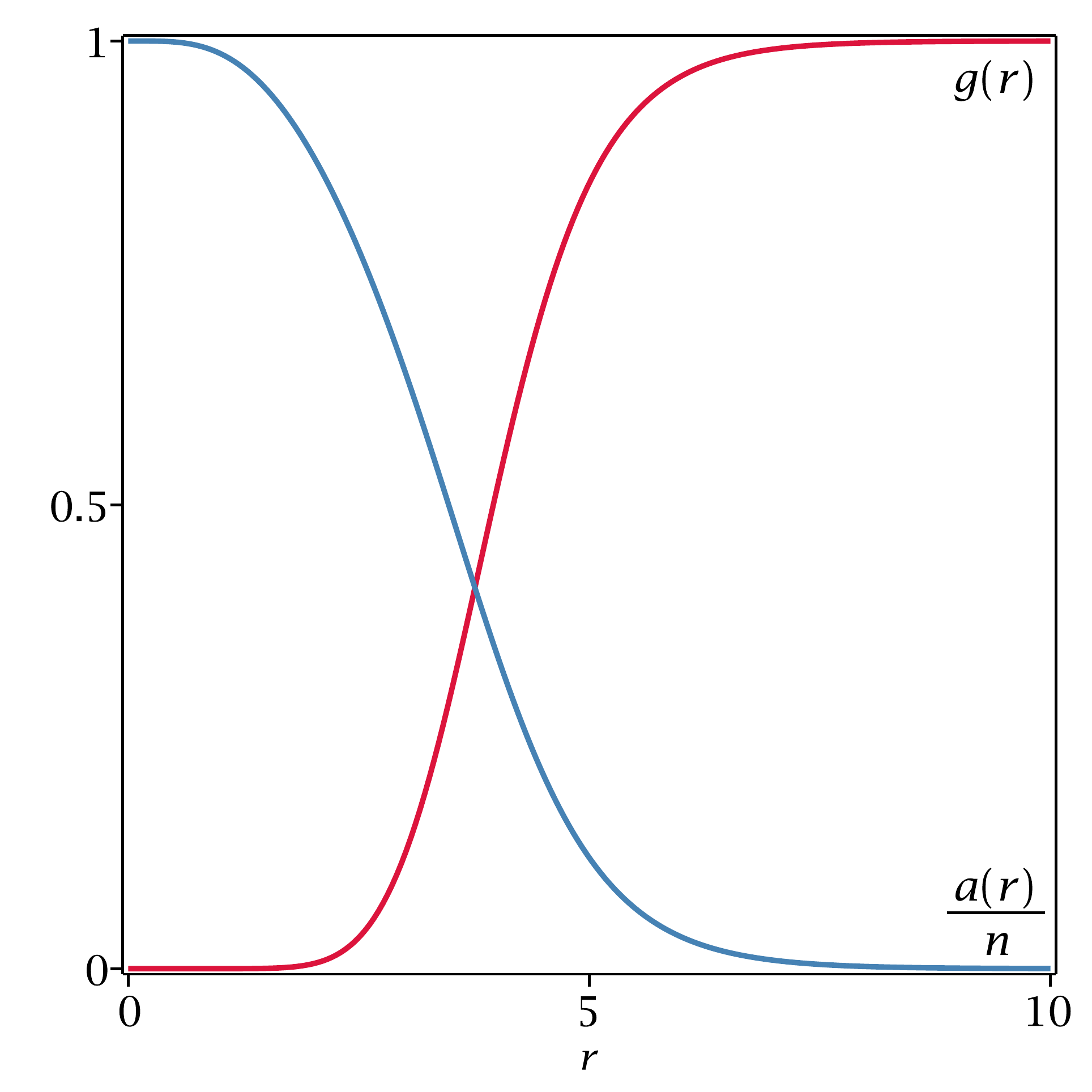}
		\includegraphics[width=4.2cm]{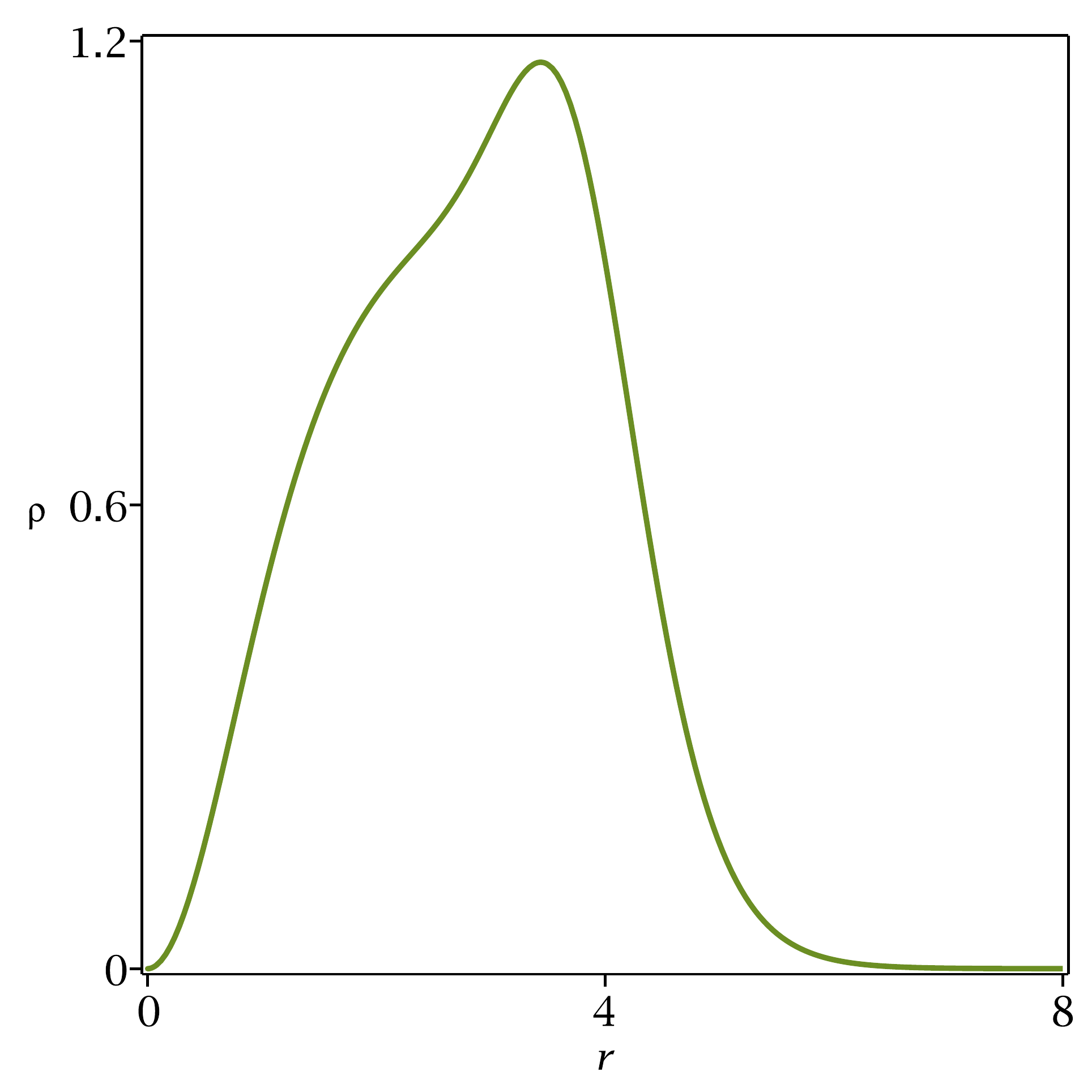}
		\includegraphics[width=4.2cm]{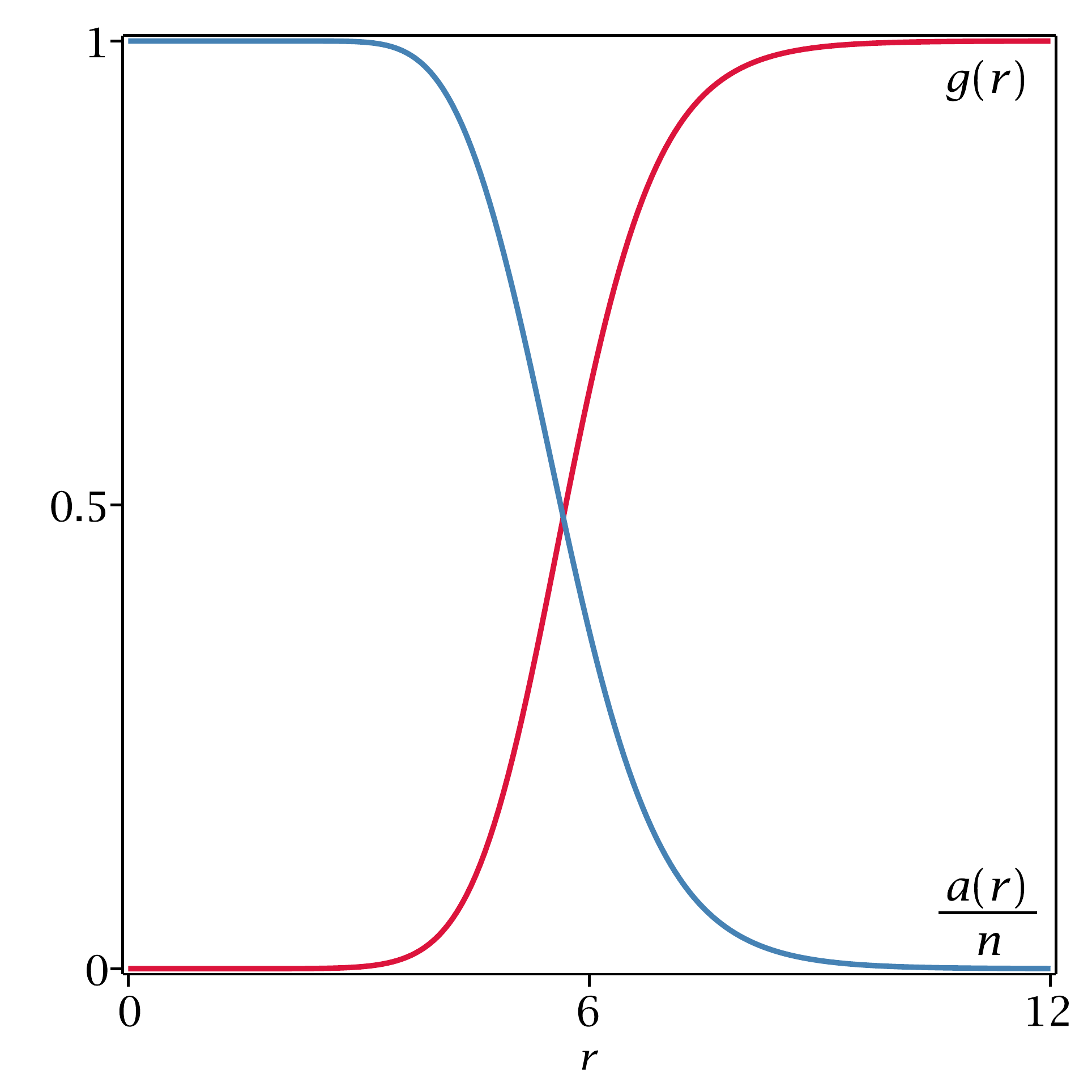}
		\includegraphics[width=4.2cm]{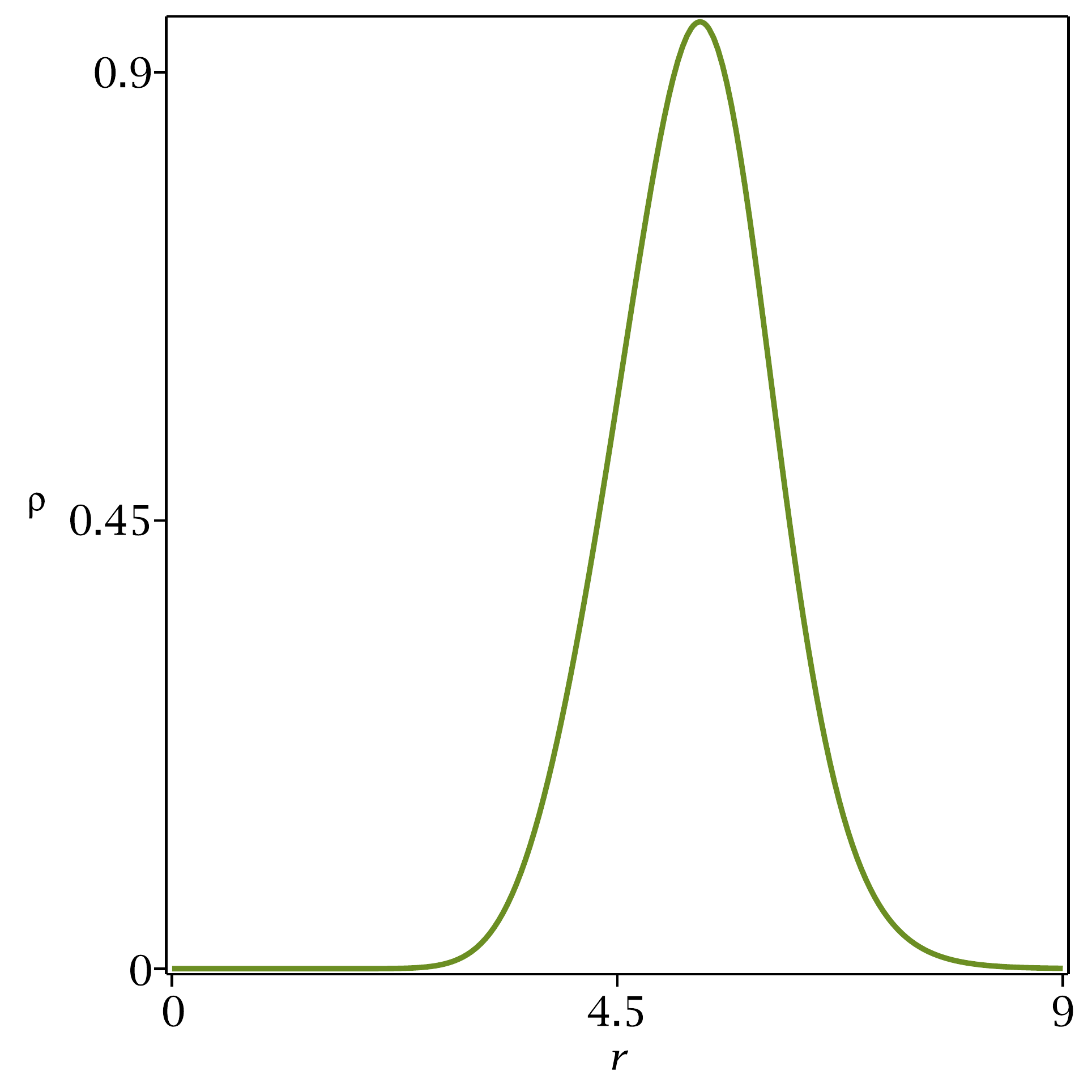}
		\caption{Solutions of~\eqref{FO} for generalized permeabilities of the form~\eqref{ex1}. In the left, we depict the vortex solutions $g_1(r),\, g_2(r),\, g_3(r)$ (red lines, from top to bottom) and $a_1(r)/n_1,\, a_2(r)/n_2,\, a_3(r)/n_3$ (blue lines, top to bottom). In the right, we depict the corresponding energy densities $\rho_1(r),\, \rho_2(r),\, \rho_3(r)$ (green lines), from top to bottom. Here, $n_1=1$, $n_2=n_3=10$.}
		\label{fig1}
	\end{figure}

\begin{figure}[h]
		\centering
		\includegraphics[width=4.8cm]{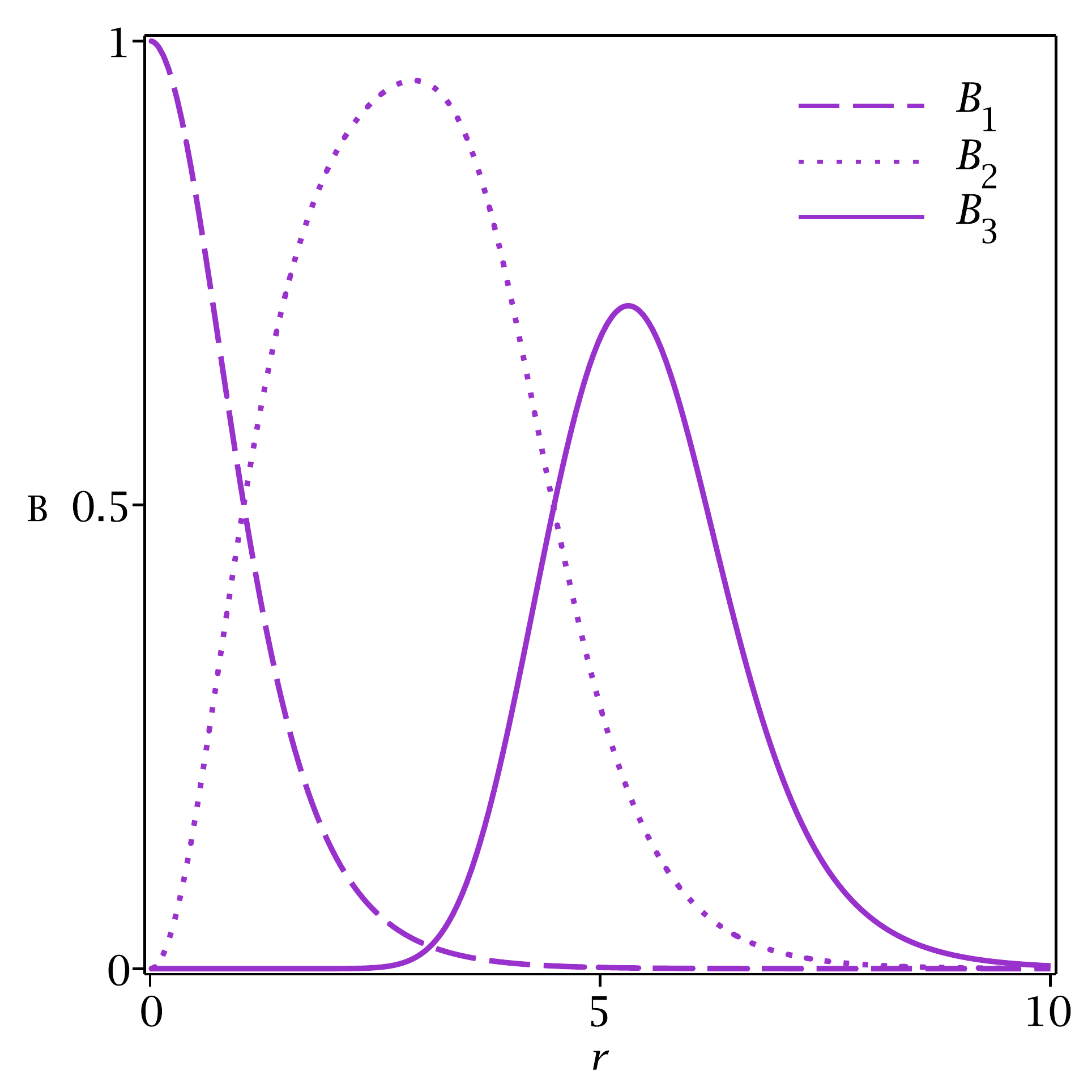}
		\caption{Magnetic inductions from sectors $1$, $2$ and $3$, for the model defined by~\eqref{ex1} with $n_1=1$, $n_2=n_3=10$.}
		\label{ex1ind}
	\end{figure}

The Bogomol'nyi equations have been solved numerically, and the results are presented bellow in Fig.~\ref{fig1}. The results were achieved through the standard procedure described, for example, in Ref.~\cite{Jackiw}, which requires solving the equations of motion in a neighborhood of the origin. This determines the solution up to a multiplicative constant, which is adjusted to match the boundary conditions at infinity. The near-origin behavior determines initial conditions to~\eqref{FO}, which can then be solved numerically. This procedure is standard and was used in all the examples presented in this paper. At a neighborhood of the origin we have, for the choice of $n_2$ depicted in that figure, $\mu_3= \mathcal{O}(22)$. This means that $\vf{3}$ solves a GL equation of the form $D_kD_k\vf{3}=\widetilde{\lambda}(1-|\vf{3}|^2)\vf{3}$, with parameter $\widetilde{\lambda}\equiv\mu_3(0)=0$, as long as terms of the order of $r^{22}$ may be neglected. By comparison, the approximation $g_3\approx 0$ breaks down at tenth-order for this choice of $n_3$. This GL equation corresponds to a type-I superconductor with vanishingly small London penetration depth~\cite{Kittel}. The energy densities have a zero at the vortex positions, and the rate at which these functions grow out of zero is heavily dependent on the winding numbers. In particular, they are greatly affected by the density of vortices from the \textit{previous} sector at the origin. Also noteworthy is the skewness of $\rho_2$  and $\rho_3$. Rather than symmetric bell-shaped profiles, they present an asymmetric character that is a reflection of the presence of vortices from other sectors at the \qt{left} of their peaks (i.e., at a smaller r). Should $\varphi_1$ have a zero at a different value of $r$, $\rho_2$ would change to reflect the new configuration. In this way, the energy densities also carry information about the position of the vortices. In Fig.~\ref{ex1ind}, the induction fields $B_a$ can be seen. As the magnetic induction of  NO vortex, $B_1$ is peaked around the origin. $B_2$ and $B_3$ on the other hand are bell-shaped, and instead possess a minimum at $r=0$.

\begin{figure}[h]
		\centering
		\includegraphics[width=3.8cm, trim={3.8cm 3.8cm 3.8cm 3.8cm}, clip]{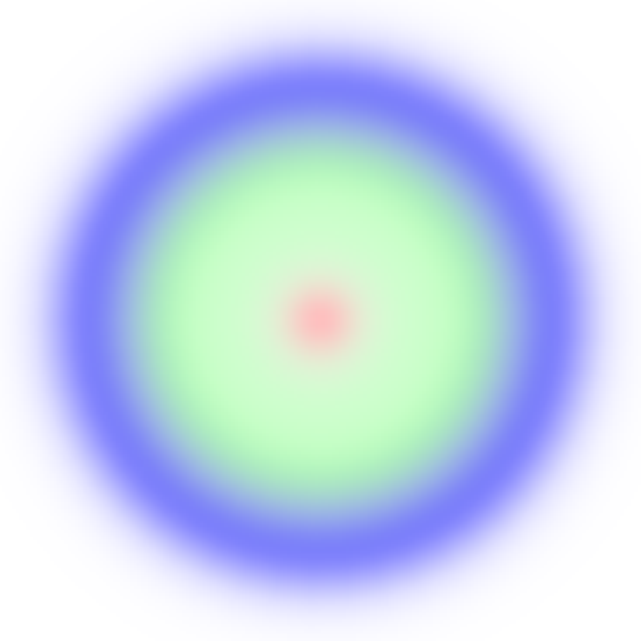}
		\includegraphics[width=3.8cm, trim={3.8cm 3.8cm 3.8cm 3.8cm}, clip]{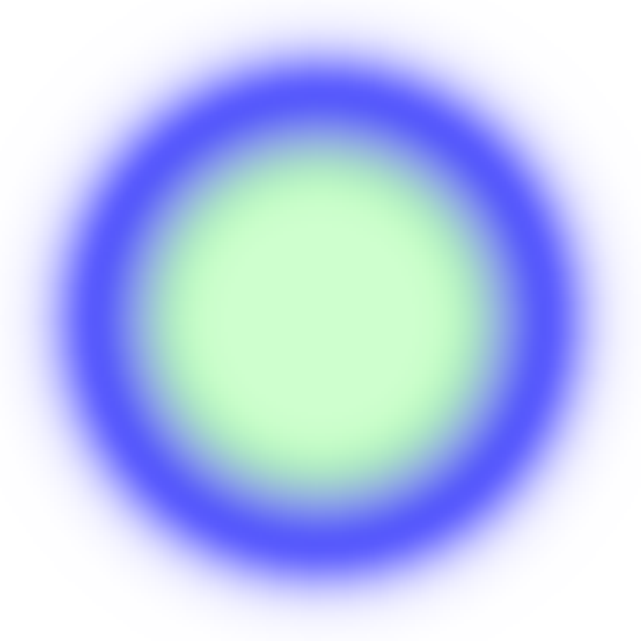}
		\caption{Planar section depicting the contributions $\rho_1$ (red), $\rho_2$ (green) and $\rho_3$ (blue) to the total energy density for solutions of~\eqref{FO} with permeabilities~\eqref{ex1}. We depict the cases $n_1=1$, $n_2=n_3=10$ (left) and  $n_1=0$, $n_2=n_3=10$ (right). The red vortex is absent from the picture on the right because $\rho_1=0$.}
		\label{mult1}
	\end{figure}

In Fig.~\ref{mult1}, planar sections of the energy densities $\rho_1$, $\rho_2$, $\rho_3$ are displayed together in the plane. Intensity of the colors in this figure is proportional to the magnitude of $\rho_b$ at that point. This construction allows us to condense a lot of information in a single picture. The radius of the shells seen in Fig.~\ref{mult1} grow with the vortex densities, since a larger $n_1$ ($n_2$) means $\rho_2$ ($\rho_3$) will grow more slowly out of their zero, and reach their peaks at a larger $r$. However, the Bogomol'nyi bound is saturated, so a given $\rho_a$ will always integrate to give the same contribution $2\pi  n_a$ to~\eqref{Energy}, meaning this vortex will have the same mass. For this to be possible, the increase in shell radius must be compensated by a decrease in the height of its peak. The emergence of shells in Fig.~\eqref{mult1} is a consequence of $\rho_a(0)=0$ which, as explained in the previous section, can only happen for $n_a>1$. A solution with $n_1=0$ is also depicted in Fig.~\ref{mult1}, in the same scale. In that case $(|\varphi_1|,|\An{1}|)$ corresponds to a trivial configuration, leaving us with only two pairs of equations in~\eqref{FO}, as in a $\rm{U(1)}^2$ model with $(\mu_2, \mu_3)=(1, |\varphi_2|^2)$. Since $E_1=0$ for a BPS solution of this sector and $\rho_1$ is nonnegative, we must have $\rho_1=0$, so that the \qt{red} vortex seem on the left of Fig.~\ref{mult1} has disappeared. Moreover, the profiles of $\rho_2$ and $\rho_3$ have also been modified, specially the former, which changed from a shell to a NO core.

\begin{figure}[h]
		\centering
		\includegraphics[width=4.2cm]{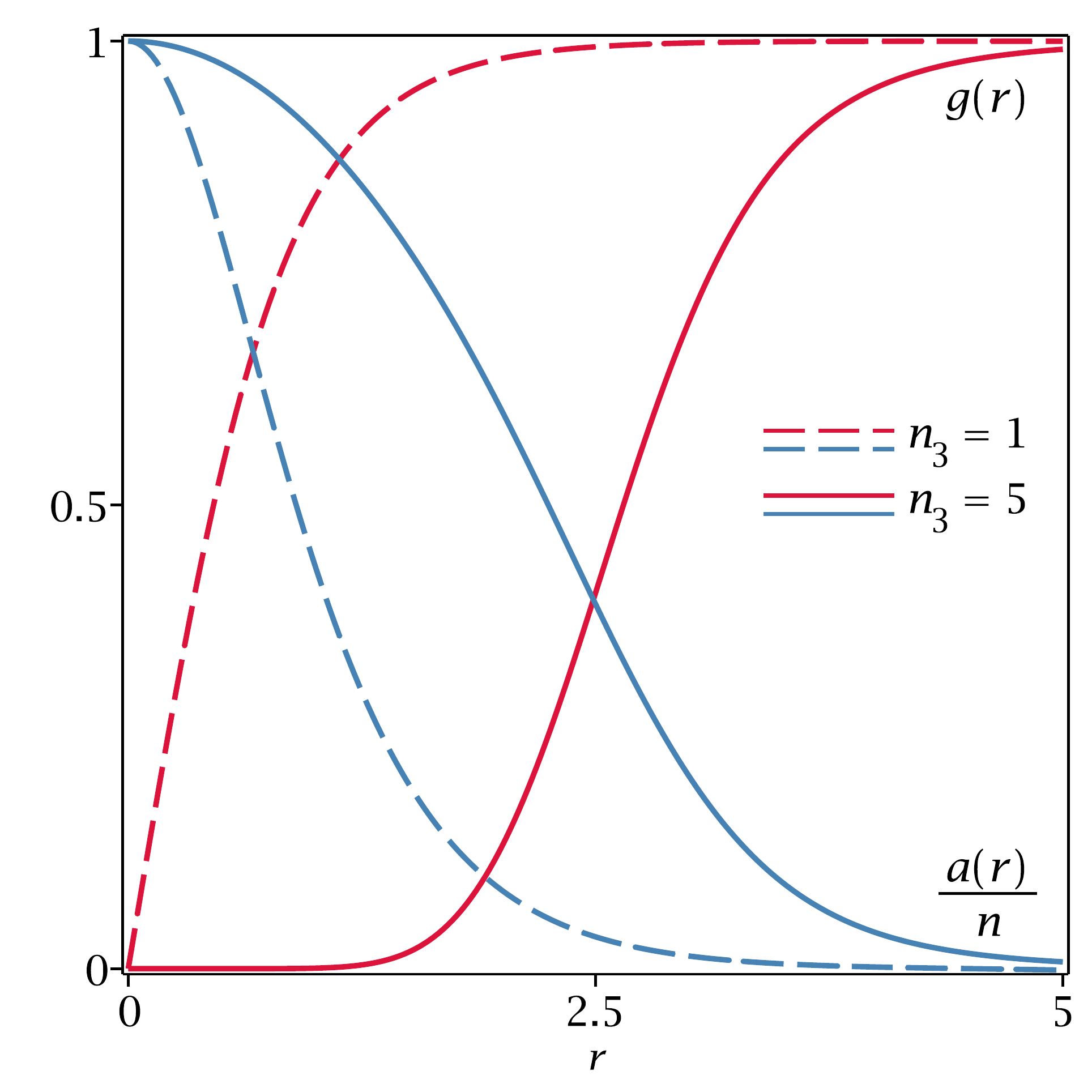}
		\includegraphics[width=4.2cm]{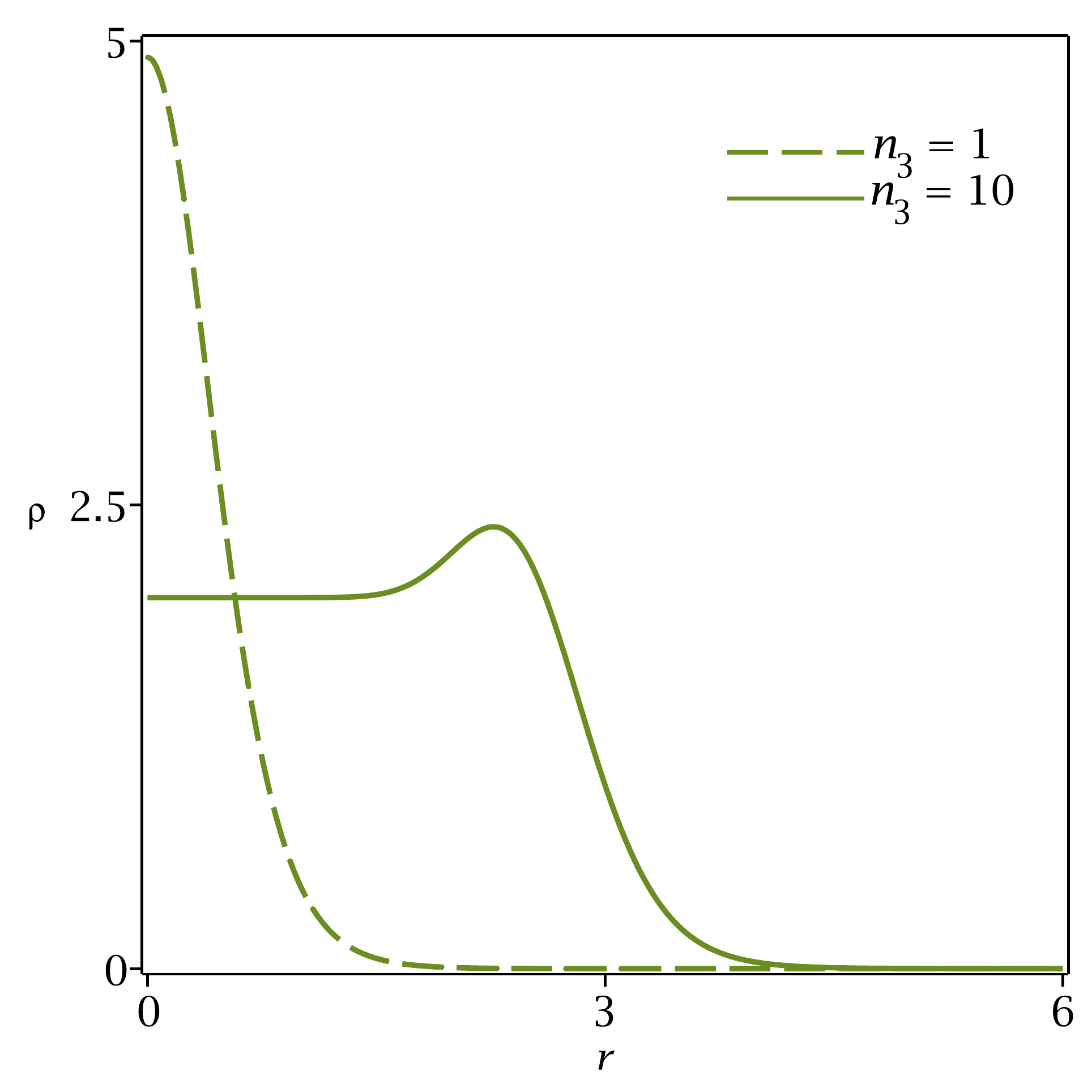}
		\caption{Solutions of~\eqref{FO} for generalized permeabilities $\mu_1=1$, $\mu_2=|\varphi_1|^2$ and $\mu_3=2-|\varphi_2|^2$. In the left, we show the solution profiles $g_3(r)$ (red) and $a_3(r)/n_3$ (blue). In the right, we depict the corresponding energy densities $\rho_3(r)$. The cases $n_3=1$ (dashed line) and $n_3=10$ (solid line) are shown.}
		\label{fig4}
	\end{figure}

\begin{figure}[h]
		\centering
		\includegraphics[width=4.8cm]{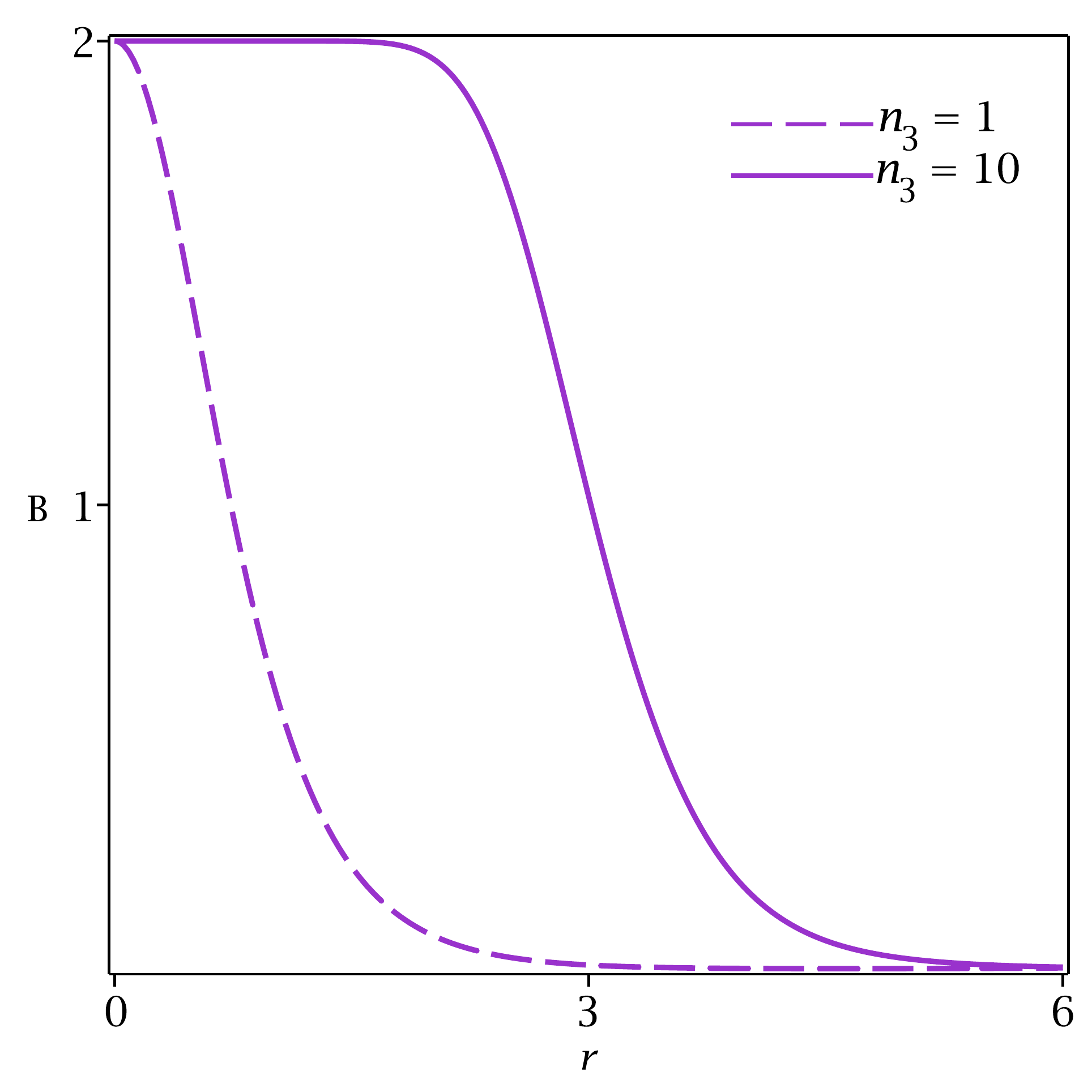}
		\caption{Magnetic inductions from sector $3$ for the model specified by~\eqref{ex1_5}. We depict the cases $n_1=1$ (solid line) and $n_2=10$ (dashed line).}
		\label{ex1_5ind}
	\end{figure}

The discussion developed in this subsection could be adapted to the case in which some of the $\mu_a$ have a maximum at the origin. To see how this works, let $\mu_1$ and $\mu_2$ be unchanged and take
	\begin{equation}\label{ex1_5}
		\mu_3=2 - |\varphi_2|^2,
	\end{equation} 
which has a maximum at the origin, where $|\varphi_2|=0$, and decreases monotonically to unit as $\varphi_2$ approaches the vacuum. Here, $\mu_3=2 - C_2r^{2{n_2}} + \mathcal{O}(2{n_2}+2)$, so $\vf{3}$ solves $D_kD_k\vf{3}=2(1-|\vf{3}|^2)\vf{3}$ in a neighborhood of the origin. Precisely as was the case for the solution depicted in Fig.~\ref{fig1}, the order of approximation up to which $\mu_3$ may be treated as a constant is determined by $n_2$. This solution is depicted in Fig.~\ref{fig4}, while the induction field is illustrated in Fig.~\ref{ex1_5ind}. The profiles of $\rho_3$ and $B_3$ are qualitatively similar to those found in the NO vortex, but the modified model allows a greater magnetic flux through the core of the defect. The $n_3=1$ case, in particular, may be directly compared to the results depicted in Figs.~\ref{fig1} and~\ref{ex1ind} for the first sector. In the modified model, the magnetic induction at the center is twice as great. As $r$ increases, this field goes to zero faster than in the standard vortex, in a way consistent with the fact they produce numerically equal fluxes, of value $2\pi$.
	
\subsection{Quadratic permeabilities and $\varphi^6$ potential}	
As the next example, let
	\begin{align}\label{ex3}
		\mu= (|\varphi_1|^2, |\varphi_1|^2, |\varphi_2|^2).
	\end{align}
$\mu_2$ and $\mu_3$ are as before, but now $\mu_1$ depends explicitly on $\vf{1}$. Thus, this vortex will not revert to a NO profile when configurations from different sectors are absent. In fact, this choice gives rise to a $|\varphi|^6$ potential of the form considered in~\cite{Jackiw} for a Chern-Simons-Higgs theory. Both $V_{1}$ and $V_{2}$ are zero when $|\varphi_1|=0$, but unlike the standard $|\varphi|^6$ model, no vacuum state is associated with this value, so a symmetric minimum of the energy does not exist. The solution of~\eqref{FO} with permeabilities~\eqref{ex3} is depicted below, in Fig. \ref{fig5} along with the $\rho_a$. Also, in Fig. \ref{ex3ind} we display the corresponding magnetic inductions. For a better view of the magnetic structure, in  Fig.~\ref{mult3}, we show a planar section of the energy densities, where a multi-shell structure can be now be seen, owing to the fact that $\rho_1$ now has itself a zero at the origin. As in Fig.~\ref{mult1}, one could set $n_1=0$ to turn the green shell into a core. Removal of the green shell, achieved by setting $n_2=0$, would turn the blue shell into a core centered at the origin, while leaving the red shell unchanged.
	\begin{figure}[h]
		\centering
		\includegraphics[width=4.2cm]{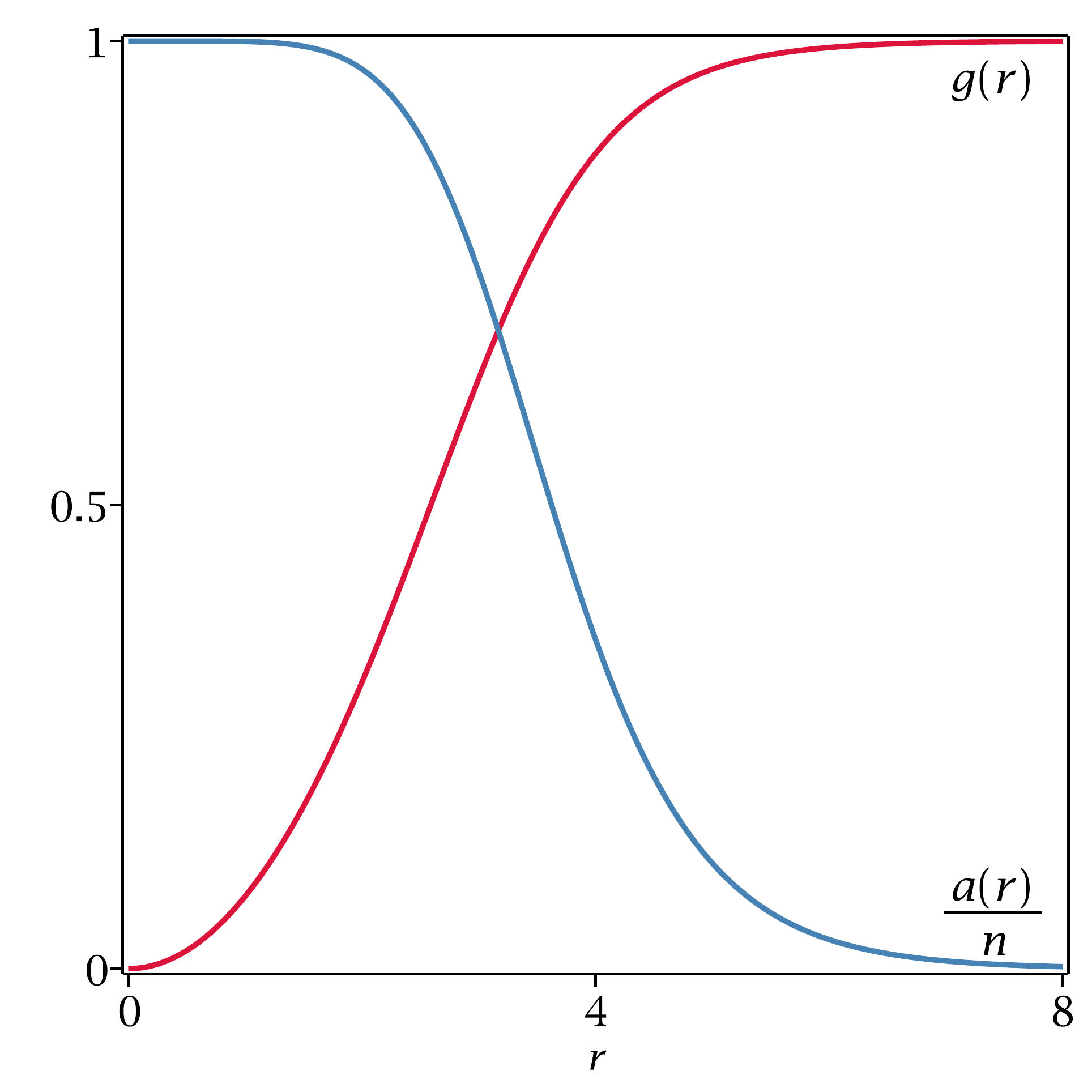}
		\includegraphics[width=4.2cm]{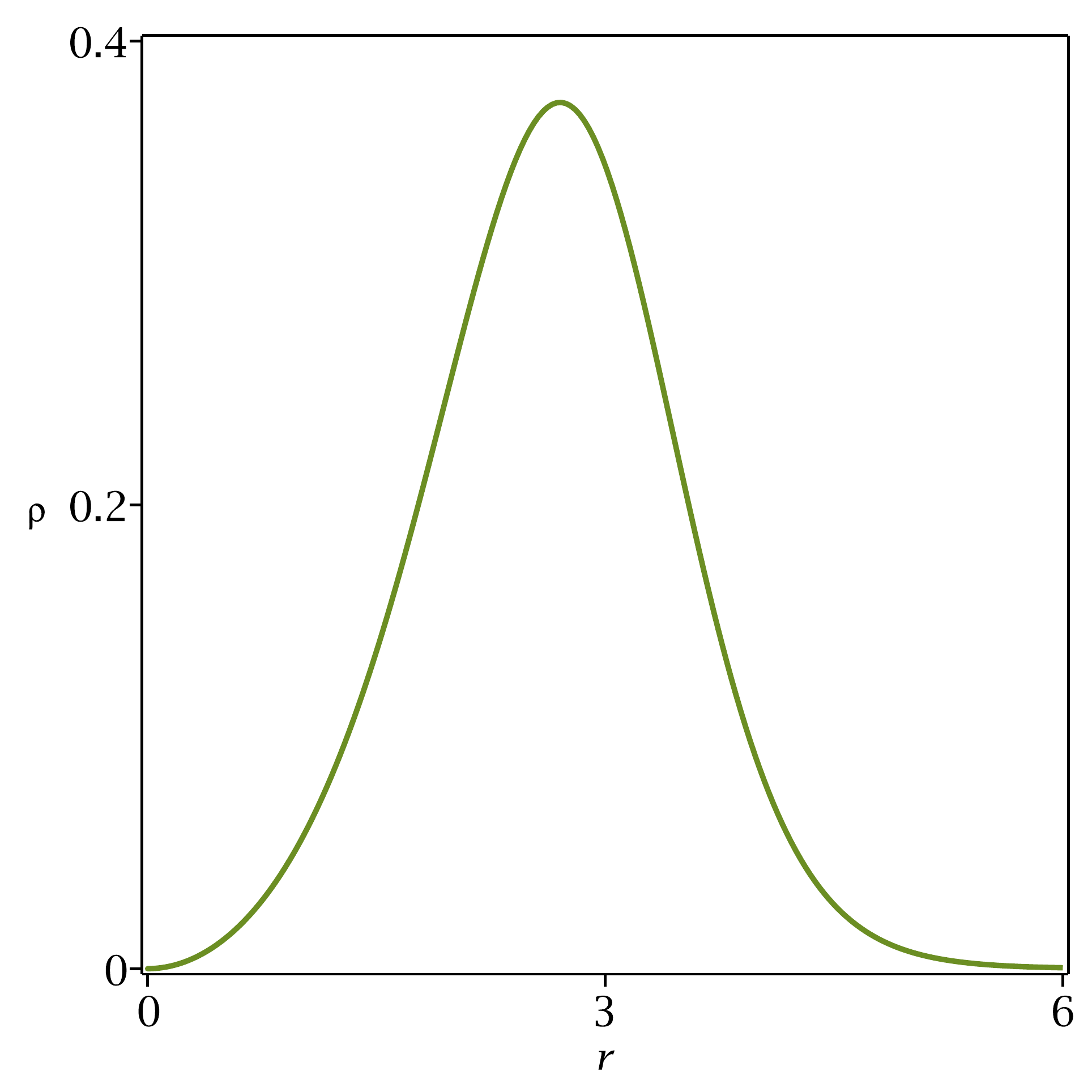}
		\includegraphics[width=4.2cm]{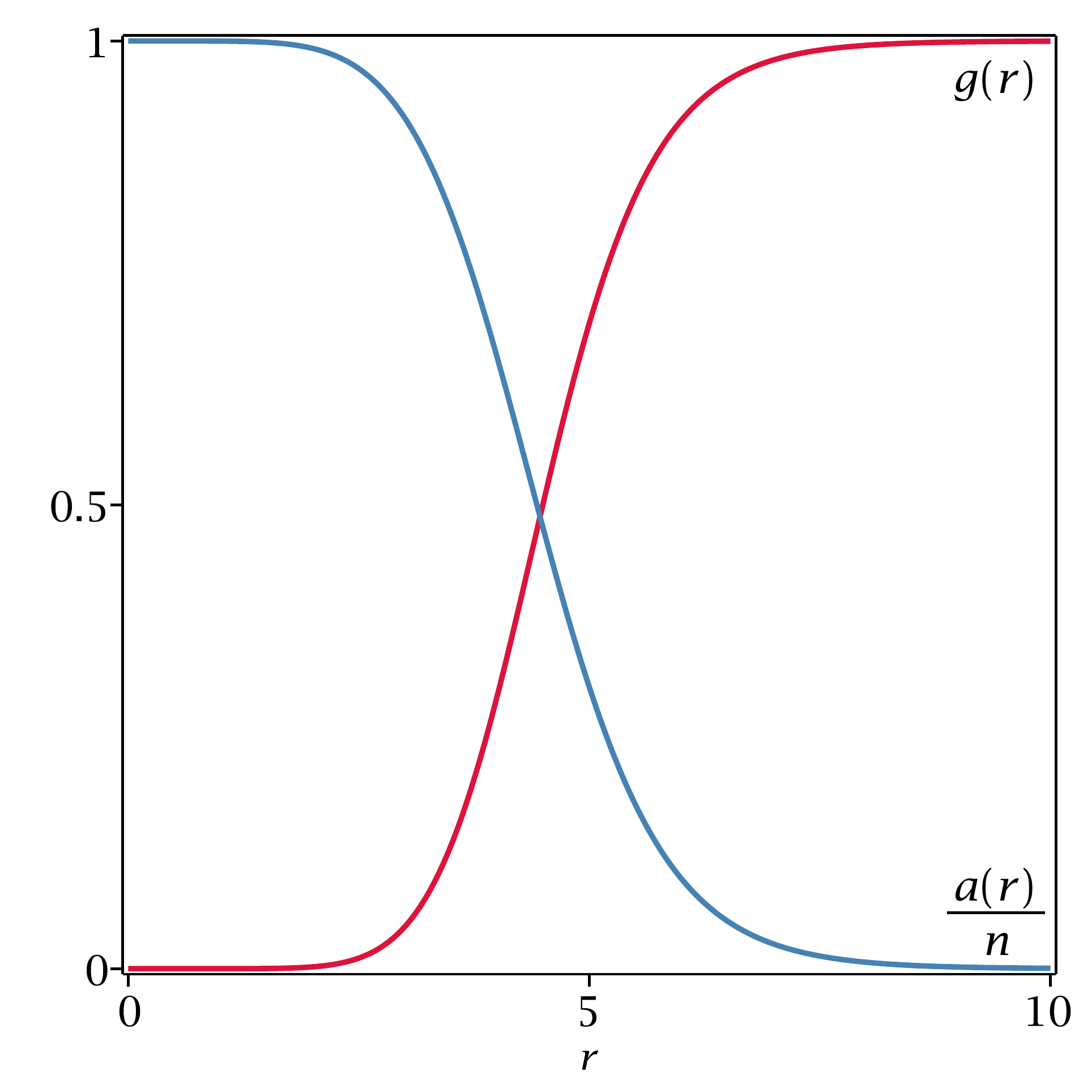}
		\includegraphics[width=4.2cm]{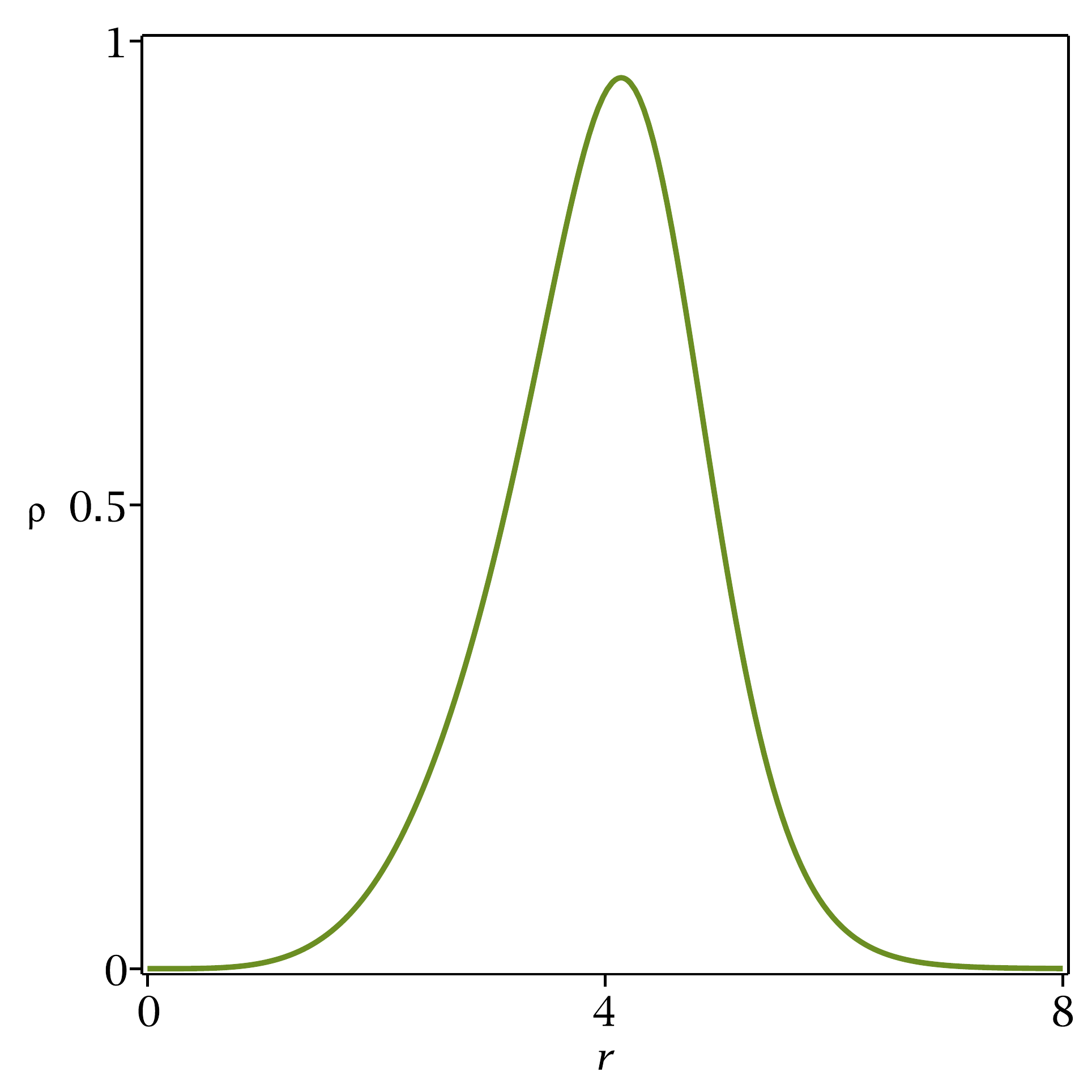}
		\includegraphics[width=4.2cm]{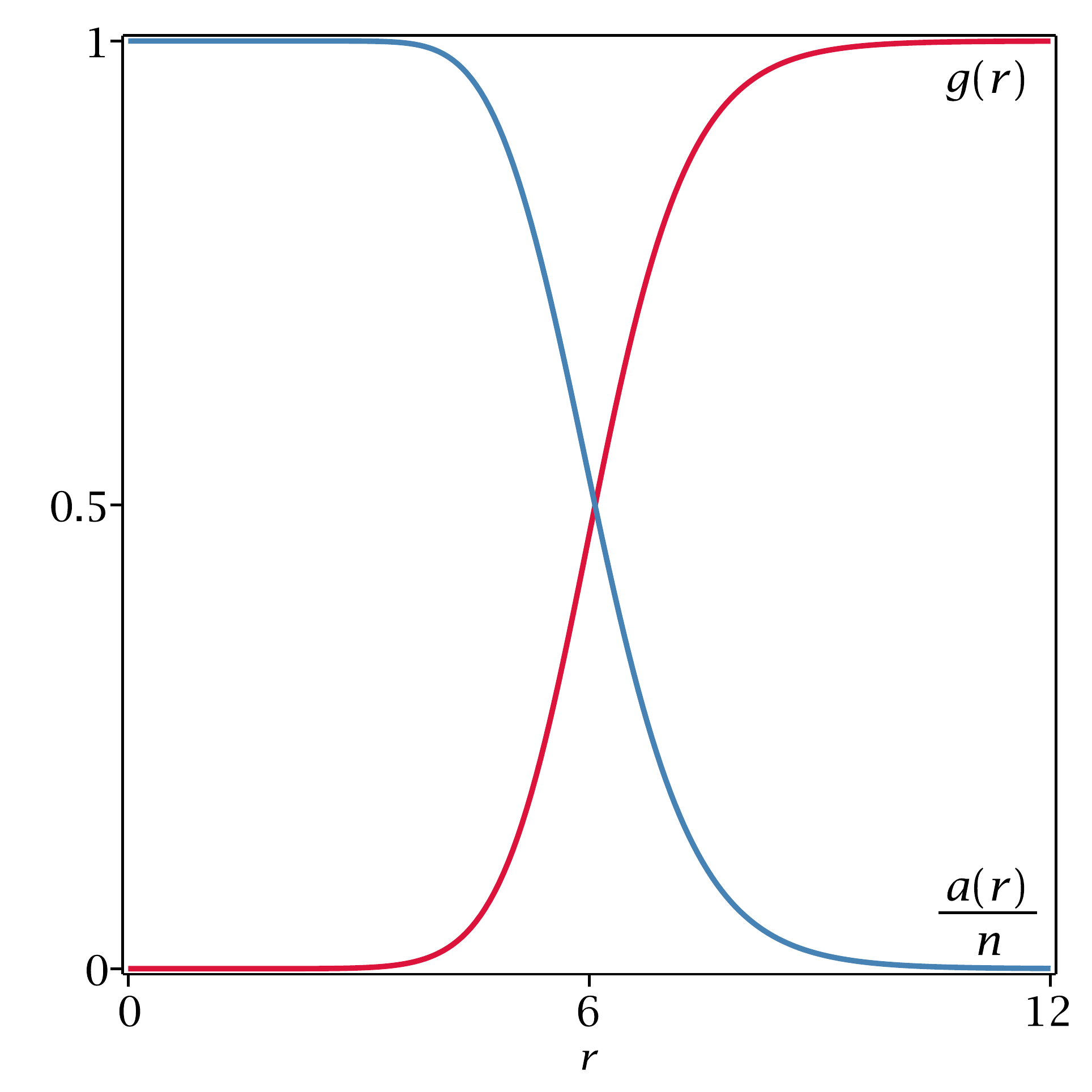}
		\includegraphics[width=4.2cm]{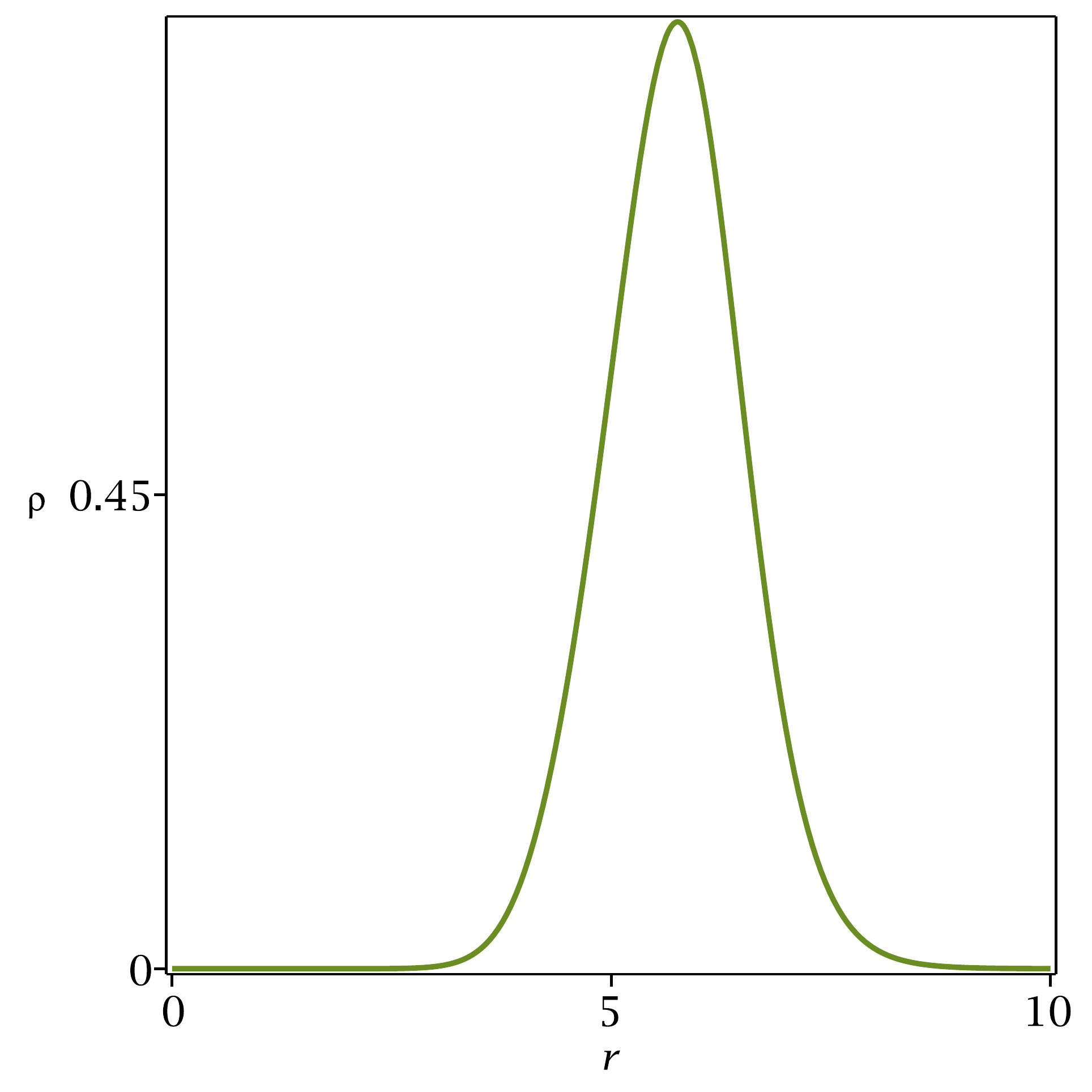}
		\caption{Solutions of~\eqref{FO} with permeabilities given by~\eqref{ex3}. In the left, we depict the vortex solutions $g_1(r),\, g_2(r),\, g_3(r)$ (red lines, from top to bottom) and $a_1(r)/n_1,\, a_2(r)/n_2,\, a_3(r)/n_3$ (blue lines, from top to bottom). In the right, we depict the corresponding energy densities $\rho_1(r), \rho_2(r)$, and $\rho_3(r)$ (top to bottom). Here, $n_1=2$, $n_2=8, \, n_3=10$.}
		\label{fig5}
	\end{figure}
	\begin{figure}[h]
		\centering
		\includegraphics[width=4.8cm]{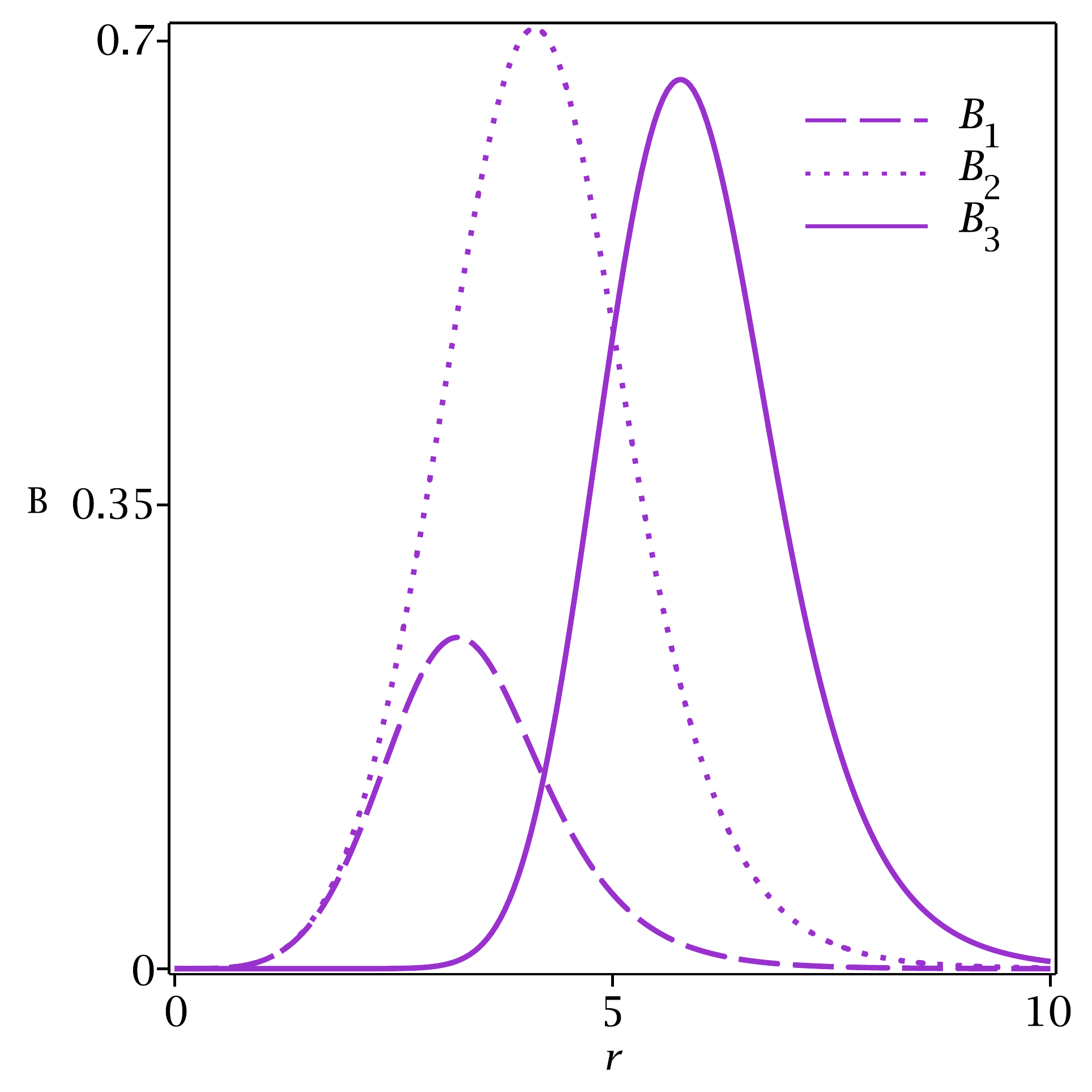}
		\caption{Magnetic inductions from sectors $1$, $2$ and $3$, with $n_1=2$, $n_2=8$, $n_3=10$ and permeabilities given by~\eqref{ex3}.}
		\label{ex3ind}
	\end{figure}
	\begin{figure}[h]
		\centering
		\includegraphics[width=4cm, trim={4.4cm 4.4cm 4.4cm 4.4cm}, clip]{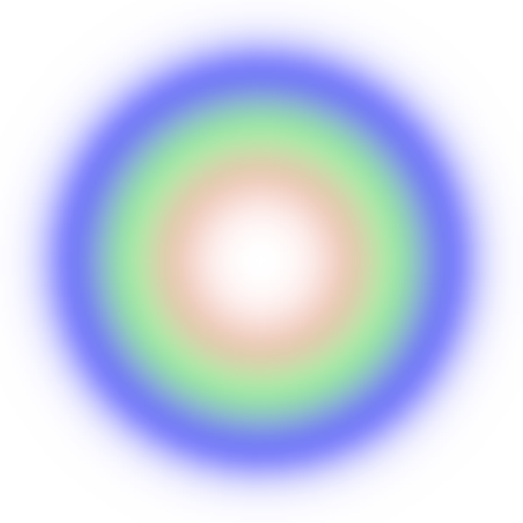}
		\caption{Planar section depicting the contributions $\rho_1$ (red), $\rho_2$ (green) and $\rho_3$ (blue) to the total energy density for solutions  of~\eqref{FO} with permeabilities given by~\eqref{ex3}. We depict the case $n_1=2$, $n_2=8$, $n_3=10$. }
		\label{mult3}
	\end{figure}

The vortex $(g_1, a_1)$ gives rise to both the same induction and scalar fields as a symmetric Chern-Simons solution. These solutions are however very different form a physical point of view, as the latter is charged, and does not give rise to fields that obey Maxwell's equations. If $n_1=n_2$ in this example, it is found that the  BPS equations for $g_2$ and $a_2$ are solved by the solution $(g_1, a_1)$ and, therefore, gives rise to a magnetic induction mathematically equivalent to that of a Chern-Simons-Higgs solution. By extension, the same reasoning applies to $g_3$ when $n_1=n_2=n_3$. An interesting feature is that, when the red core from Fig.~\eqref{mult3} is removed (as would happen, for example, if the corresponding vortex is scattered away to infinity), $(g_2, a_2)$ changes from a Chern-Simons-like vortex to a NO one, thus relating these two types of defects. This effect was identified long ago in Refs. \cite{AA1,AA2}. In the present context, the defect is still uncharged and therefore the fields $E_k^{\{2\}}$ and $E_k^{\{3\}}$ are absent, but they may be added with the help of electric impurities of the form introduced by Tong and Wong in~\cite{Impurities}.

\subsection{Sinusoidal models}
\begin{figure}[h]
		\centering
		\includegraphics[width=4.2cm]{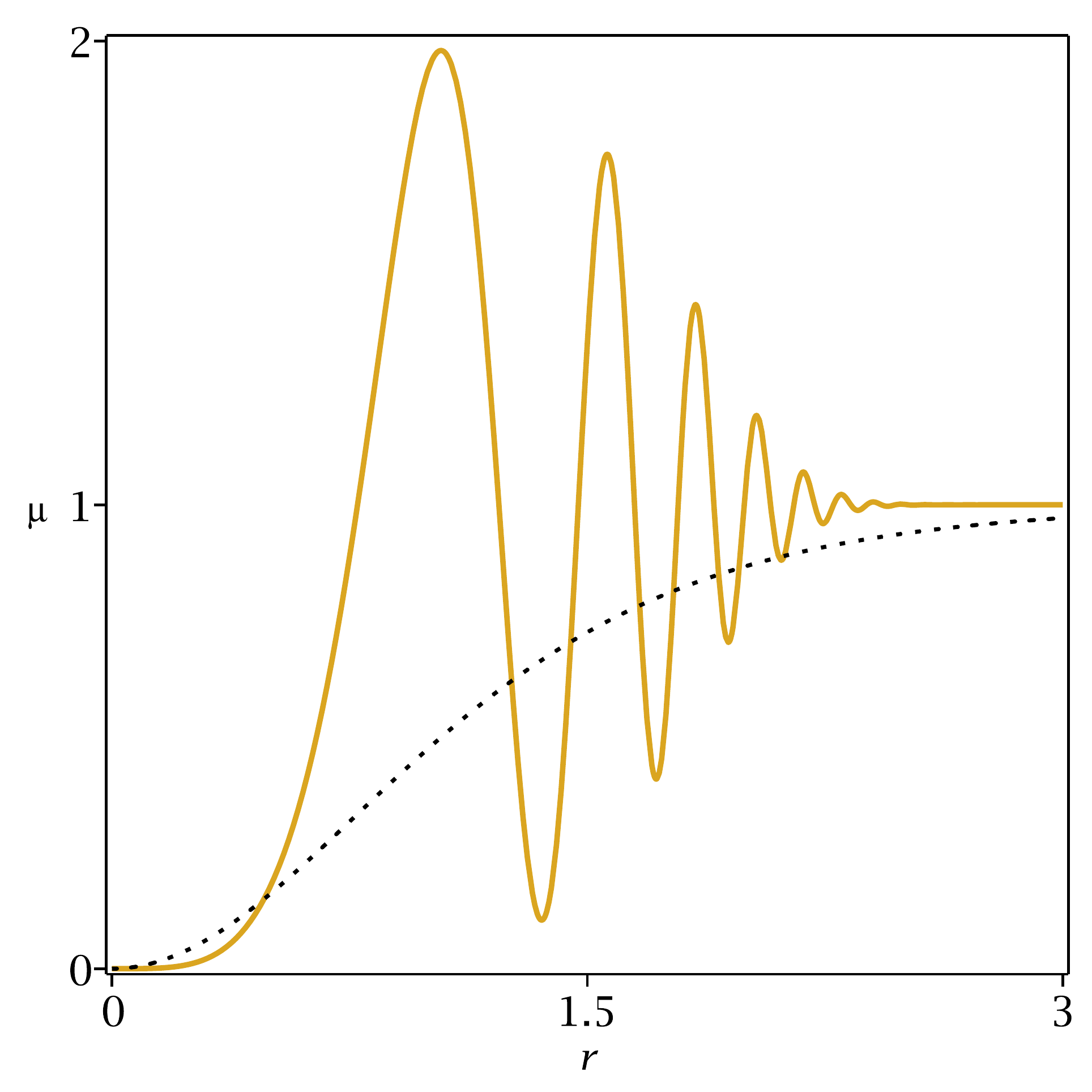}
		\includegraphics[width=4.2cm]{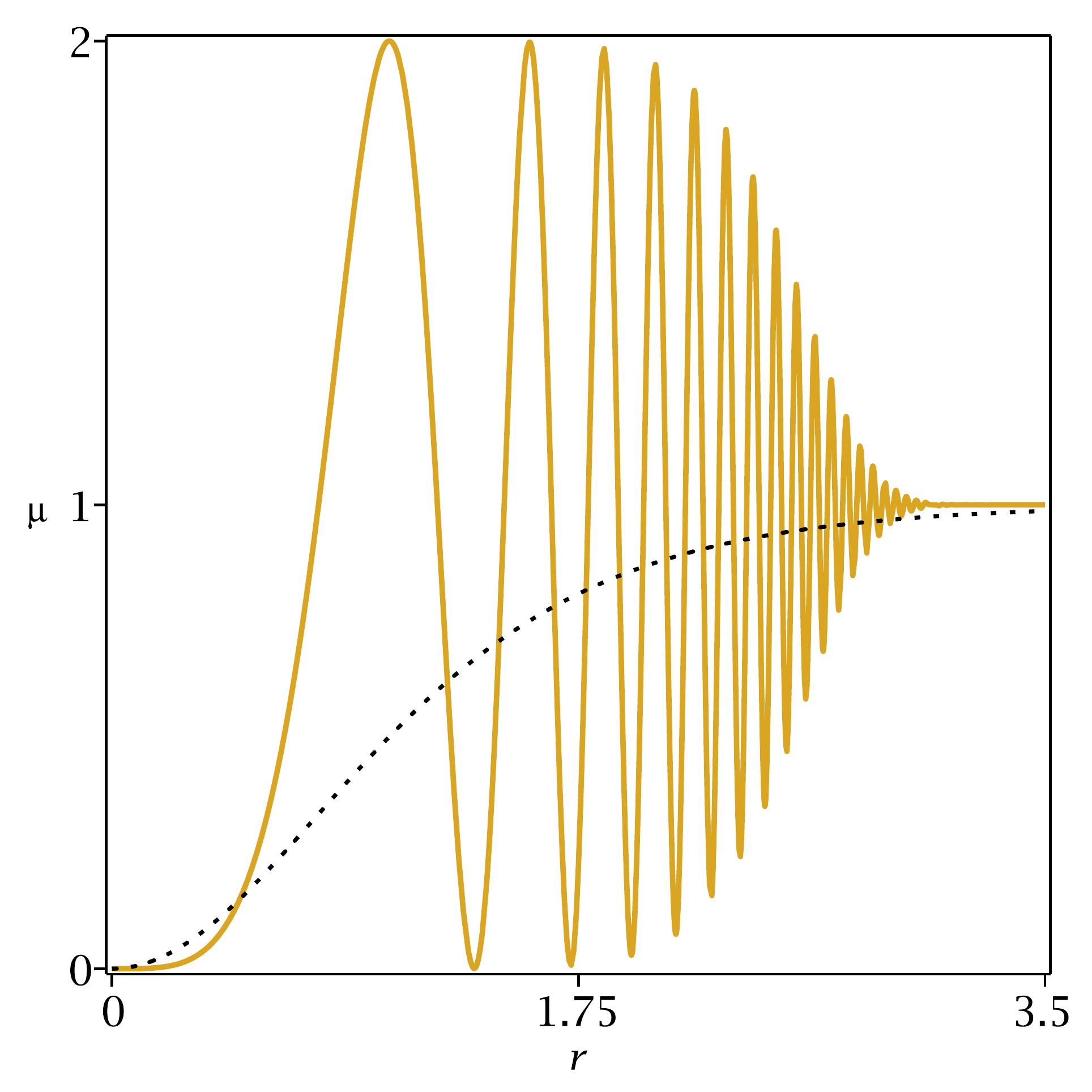}
		\caption{Permeability $\mu_3(r)$ (golden line) for $n_1=1$ and $n_2$ equal to 5 (left) and 10 (right). For comparison, the choice $\mu_3=|\vf{1}|^2$ is shown as a dotted line.}
		\label{fig6}
	\end{figure}

Next, let $\xi\equiv (1 - |\vf{1}|^2)^{-1}$ and
	\begin{align}\label{ex2}
		\mu= \left(1, |\varphi_1|^2,1 + \cos\left(\pi\xi\right)e^{-|\vf{2}|\xi} \right),
	\end{align}
for $|\vf{1}|\neq 1$ and $\mu_3=1$ for $|\vf{1}|\neq 1$. The variable $\xi$ gives a measure of how close $\vf{1}$ is from its vacuum value. Here, $\mu_1$ and $\mu_2$ are as before, and $\mu_3$ has the form of a damped oscillator on the variable $\xi$. The magnitude of $\vf{2}$ plays the role of a damping coefficient, which controls how fast the amplitude of the oscillations falls to zero. In Fig.~\ref{fig6}, we depict $\mu_3(r)$ for $n_2=5$ and $n_2=10$.

We see that different choices for $n_2$ have a much more noticeable effect on the magnetic properties than in the previous examples. Now, $\mu_3$ still has an absolute minimum at the origin, but instead of increasing monotonically to one, this function oscillates about this value, with decreasing amplitude. The greater the topological charge of the second vortex, the greater the number of oscillations needed before the equilibrium value  $\mu_3=1$ is reached. Here, both $n_1$ and $n_2$ affect the size of the disk inside which $\mu_3\simeq 0$, albeit in a far more subtle manner than in the previous examples. 

\begin{figure}[h]
		\centering
		\includegraphics[width=4.2cm]{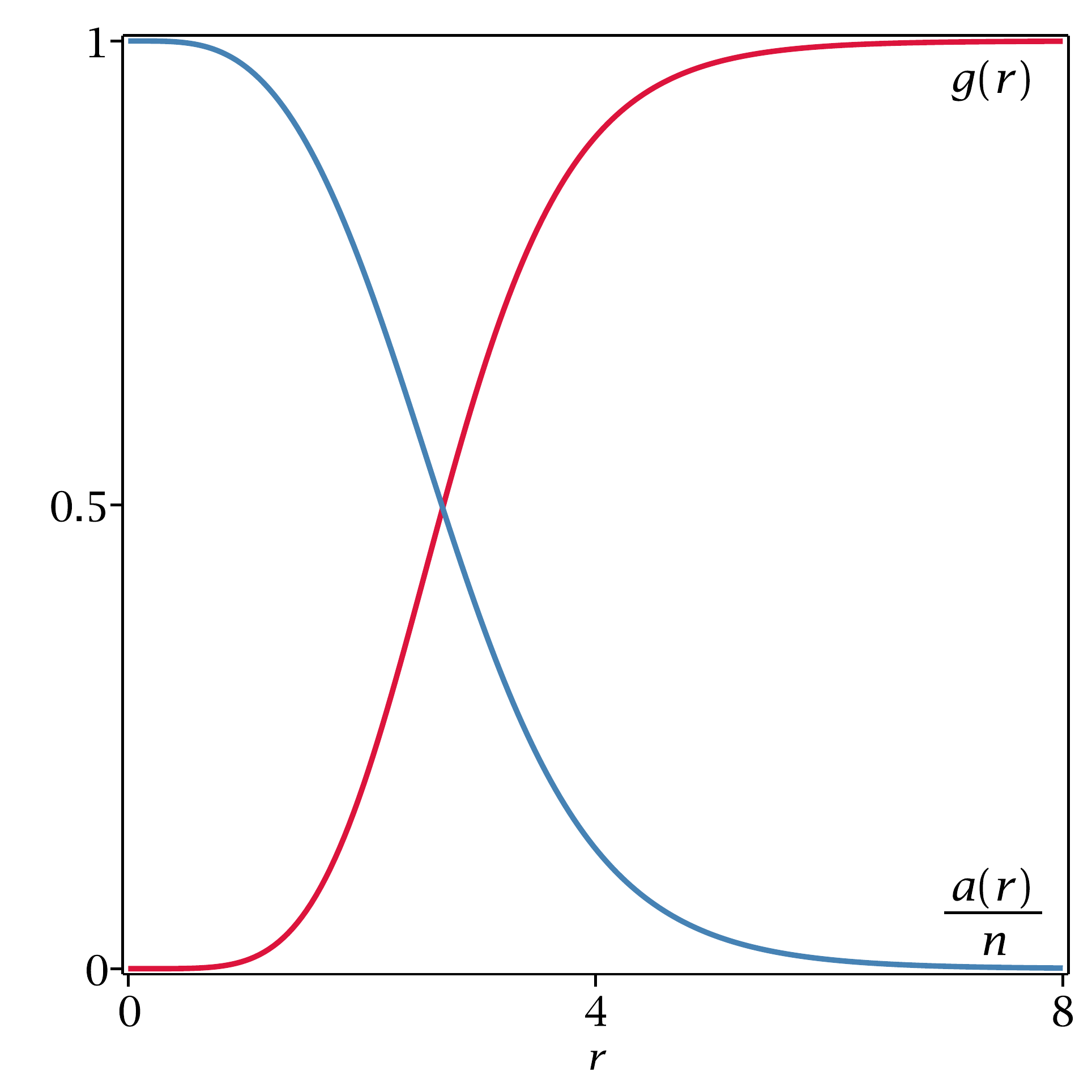}
		\includegraphics[width=4.2cm]{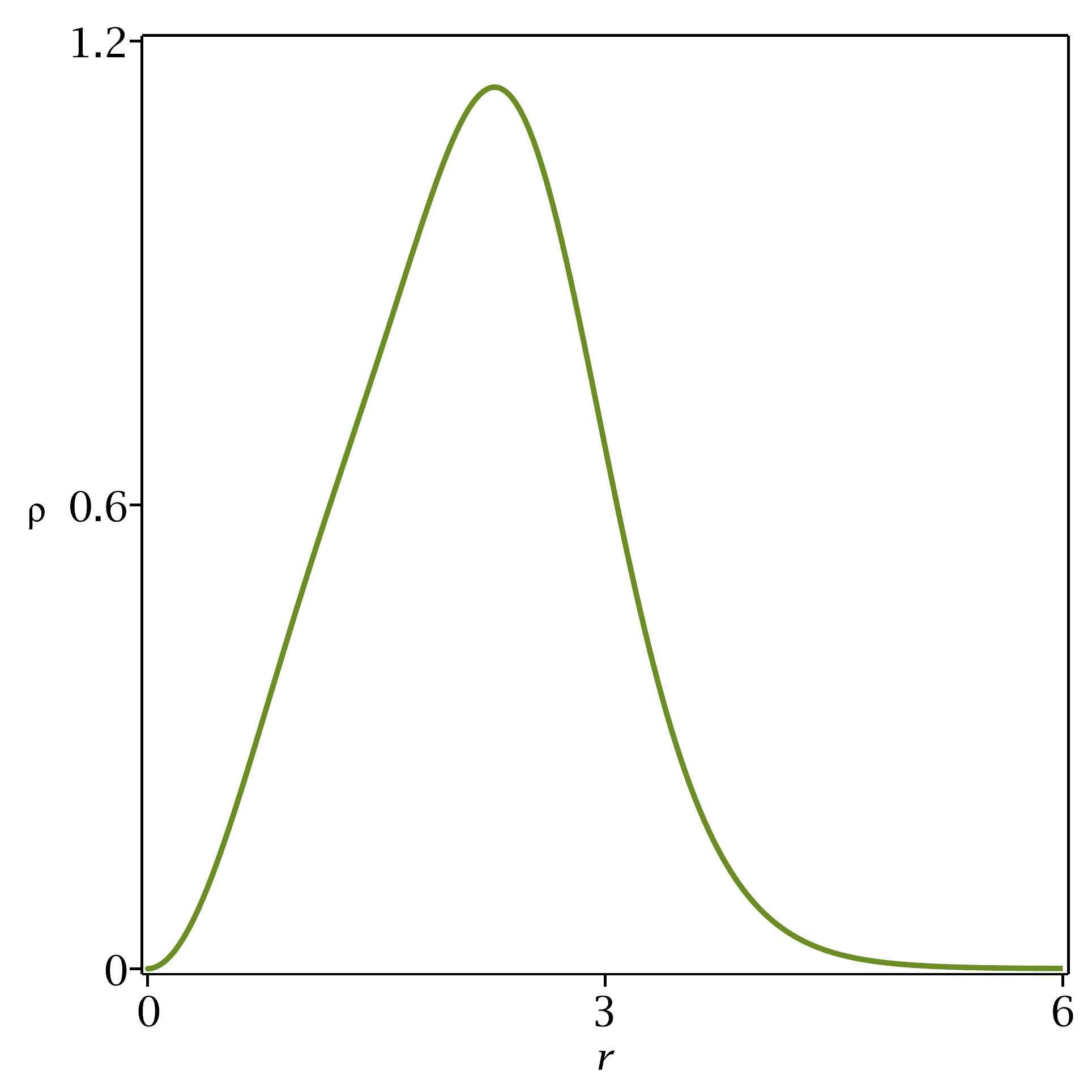}
		\includegraphics[width=4.2cm]{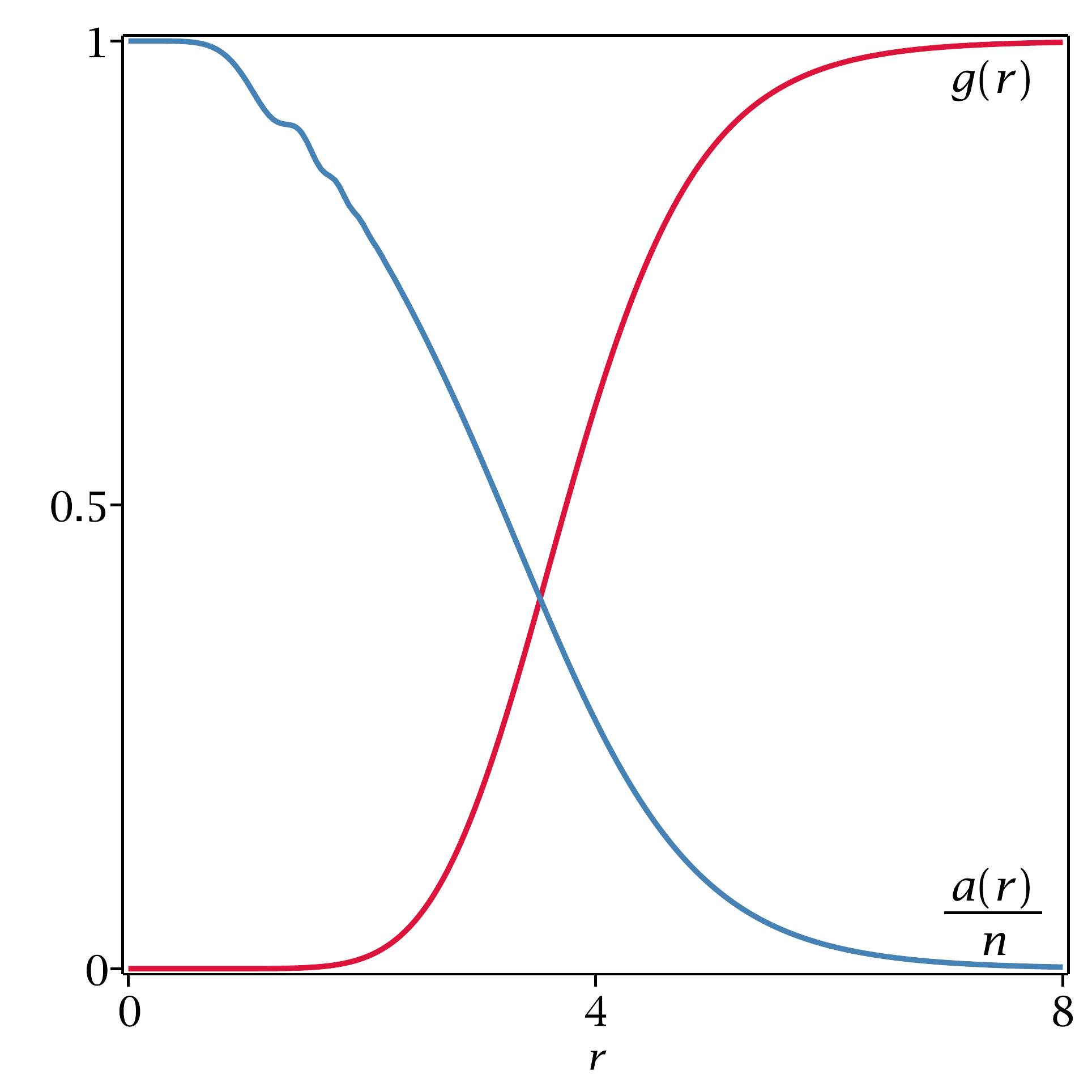}
		\includegraphics[width=4.2cm]{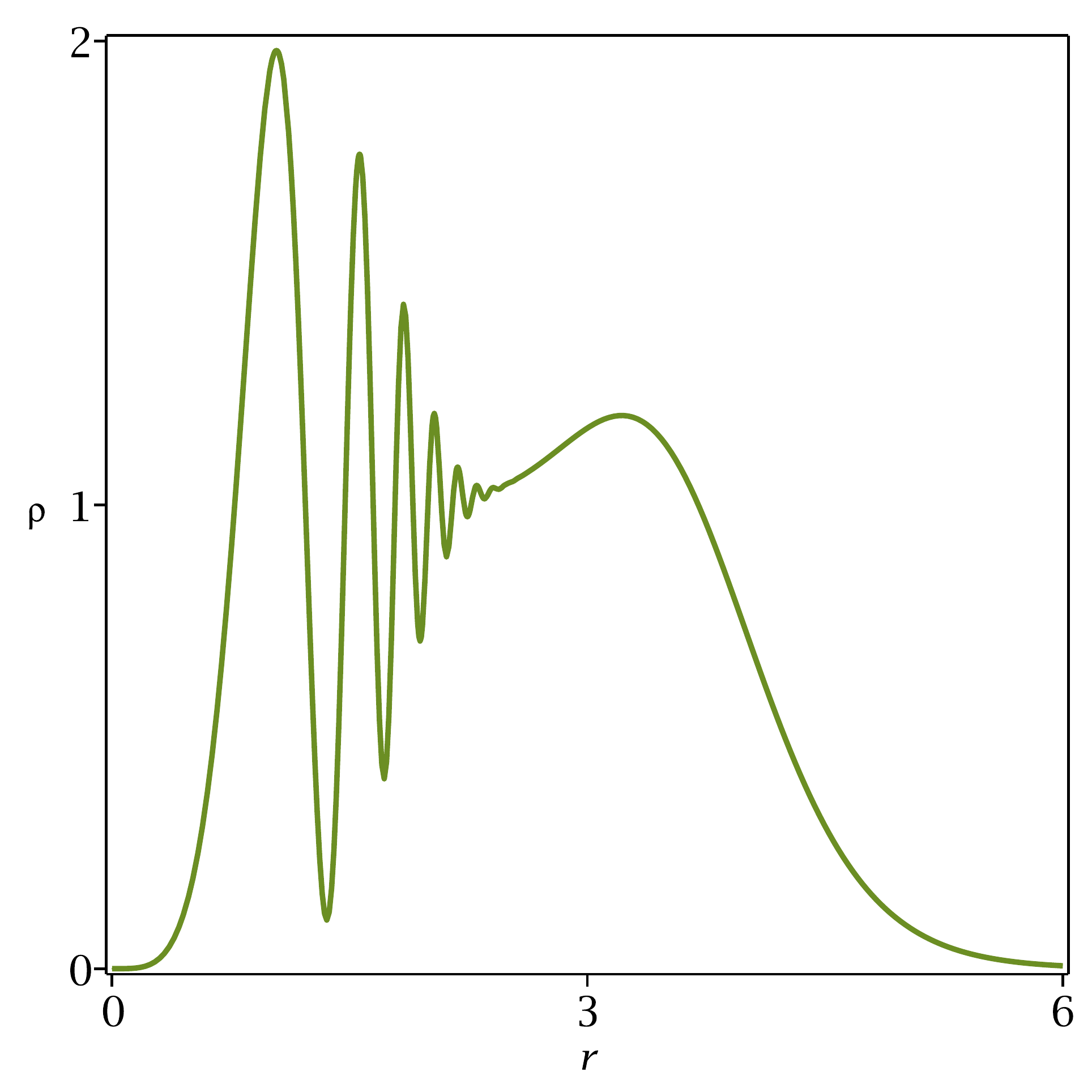}
		\caption{Solutions of~\eqref{FO} with permeabilities given by~\eqref{ex2}. In the left, we show  $g_2(r), g_3(r)$ (red lines, from top to bottom) and $a_2(r)/n_2, a_3(r)/n_3$ (blue lines, from top to bottom) for $n_1=1$, $n_2=5, n_3=10$. In the right, $\rho_2(r)$ (top) and $\rho_3(r)$ (bottom) are depicted.}
		\label{fig7}
	\end{figure}

\begin{figure}[h]
		\centering
		\includegraphics[width=4.8cm]{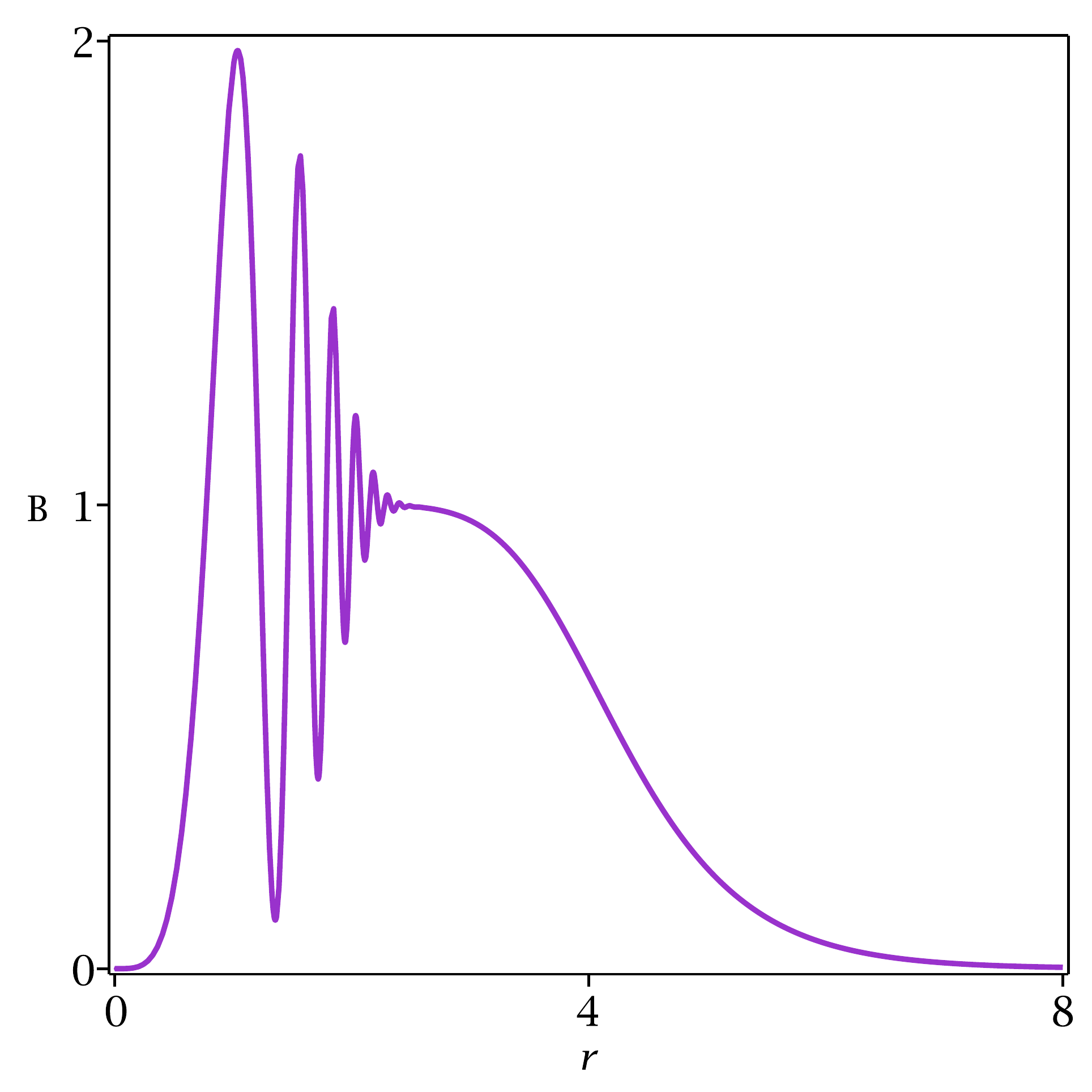}
		\caption{Magnetic induction $B_{3}(r)$, for a BPS solution  of the model specified by~\eqref{ex2}. The topological charges are $n_1=1$, $n_2=5$, and $n_3=10$. }
		\label{ex2ind}
	\end{figure}

In Fig.~\ref{fig7} we depict the fields and, in Fig.~\ref{ex2ind}, we show $B_3$ as a function of $r$, for $n_2=5$, $n_3=10$. The magnetic induction is zero at the origin and oscillates about $B_3=1$, but the damping prevents it from returning to its minimum. As $r$ grows larger, the amplitude of oscillation decreases and eventually becomes negligible when $\mu_3\simeq 1$. After this point, which depends on the winding numbers $n_1$ and $n_2$, the equations of motion involving the third sector effectively become those of a critically coupled GL theory. This happens close to $r=2.5$ for the example depicted in Figs.~\ref{fig7} and~\ref{ex2ind}. Note that all fields are still far from their vacuum values at this point, which defines a characteristic distance smaller than the size of the defects. At greater distances from the origin, $B_3$ may be treated as a monotonically decreasing function of $r$, as in a GL superconductor. These features are also expressed in the graph of $a_3$, which possesses internal structure, with a flat plateau at $r=0$ and increasingly steeper and smaller steps corresponding to local minima of $\mu_3$. As $\mu_3$ approaches unity, the internal structure disappears. 

\begin{figure}[h]
		\centering
		\includegraphics[width=4.2cm]{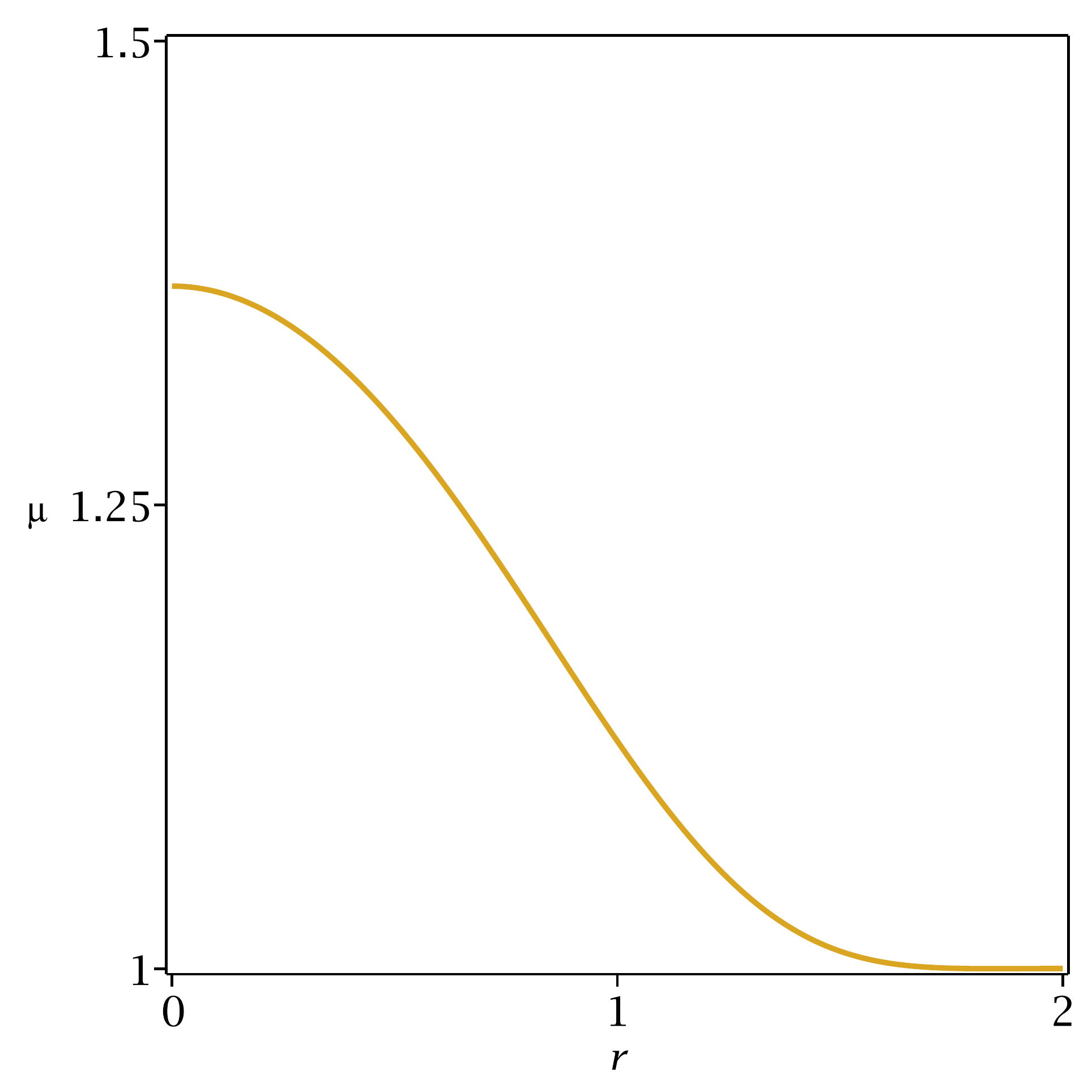}
		\includegraphics[width=4.2cm]{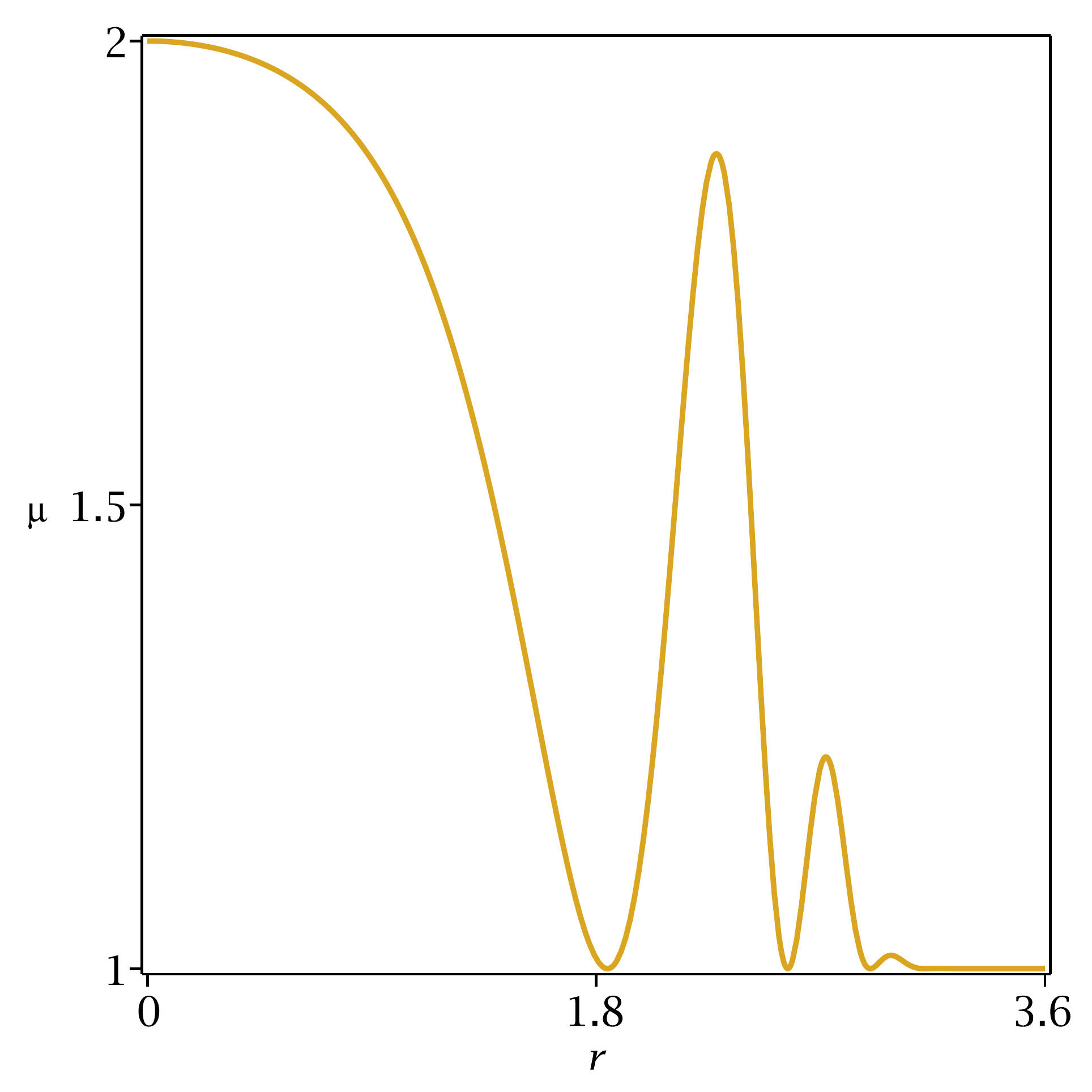}
		\caption{Permeability $\mu_3(r)$ from Eq.~\eqref{ex3_5}, for $n_1=1$. In the left, we depict $\mu_3(r)$ when $\varphi_2$ is a vacuum solution. In the right, we show the form of this permeability for a symmetric defect with $n_2=10$.}
		\label{fig9}
	\end{figure}

An interesting variation is found by changing $\mu_3$ in~\eqref{ex2} to
	\begin{equation}\label{ex3_5}
		\mu_3=1 + \alpha\cos^2\left(\beta\xi\right)e^{-|\vf{2}|\xi},
	\end{equation}
where $\alpha$ and $\beta$ are positive constants which respectively control the maximum amplitude and angular frequency of the oscillations. In this example, we take $\beta=\pi/12$ and $\alpha=\cos(\pi/12)^{-2}$, thus resulting in a maximum amplitude of one and a peak $\mu_3=2$ at the origin. In Fig.~\ref{fig9}, this choice is illustrated for $n_1=1$, and both $n_2=0$ (i.e., vacuum) and $n_2=10$ solutions of~\eqref{FO}. In the former case, the permeability presents the profile of an overdamped oscillator, which therefore falls rapidly and monotonically to unity. When, however, a non-vacuum $\vf{2}$ is present, there is a competing contribution $e^{|\varphi_2|}$ that not only slows the fall, but also eventually causes the permeability to increase, as it becomes underdamped when $|\vf{2}|$ grows. A non-vacuum $\vf{2}$ also causes the maximum of $\mu_3$ to increase in height.

\begin{figure}[h]
		\centering
		\includegraphics[width=4.2cm]{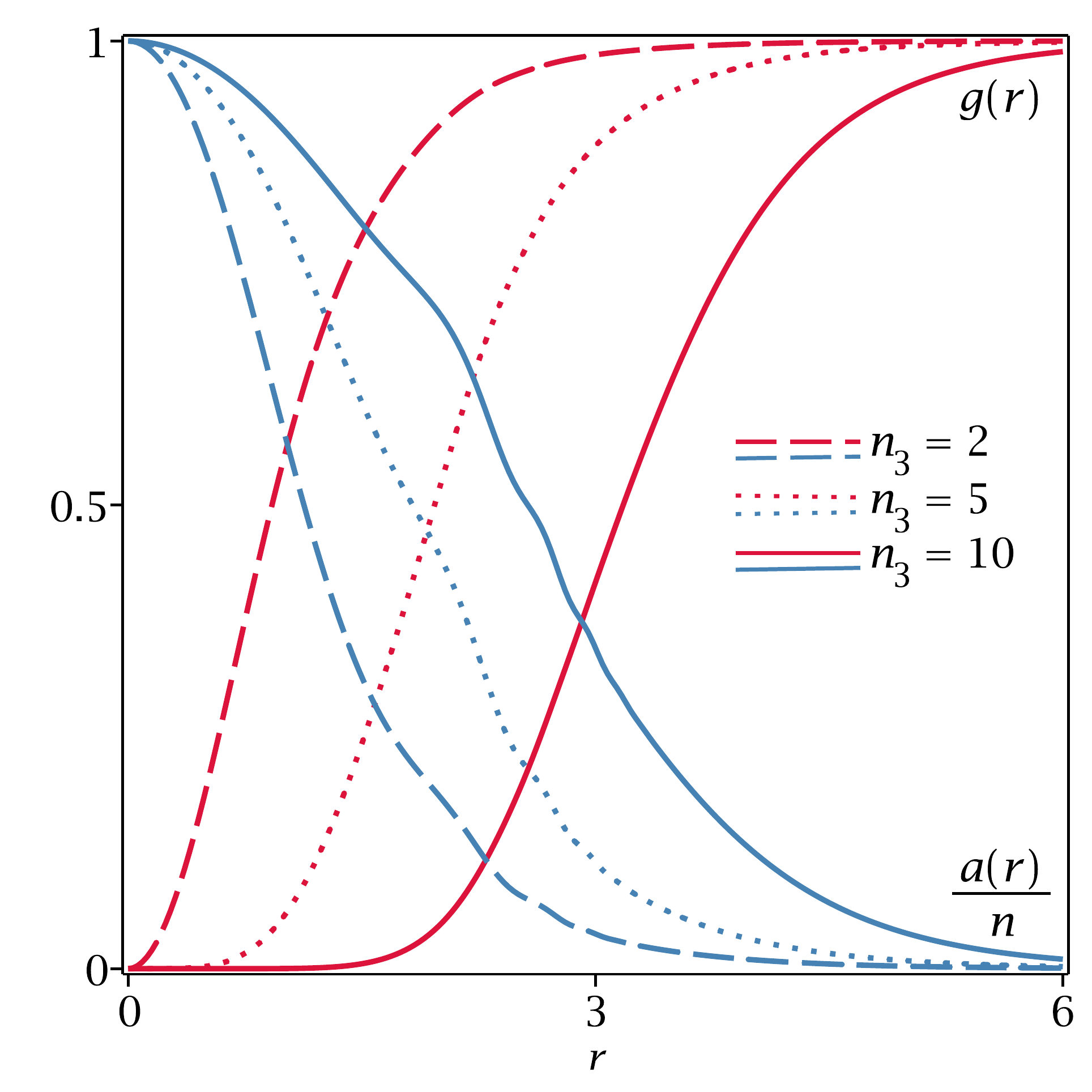}
		\includegraphics[width=4.2cm]{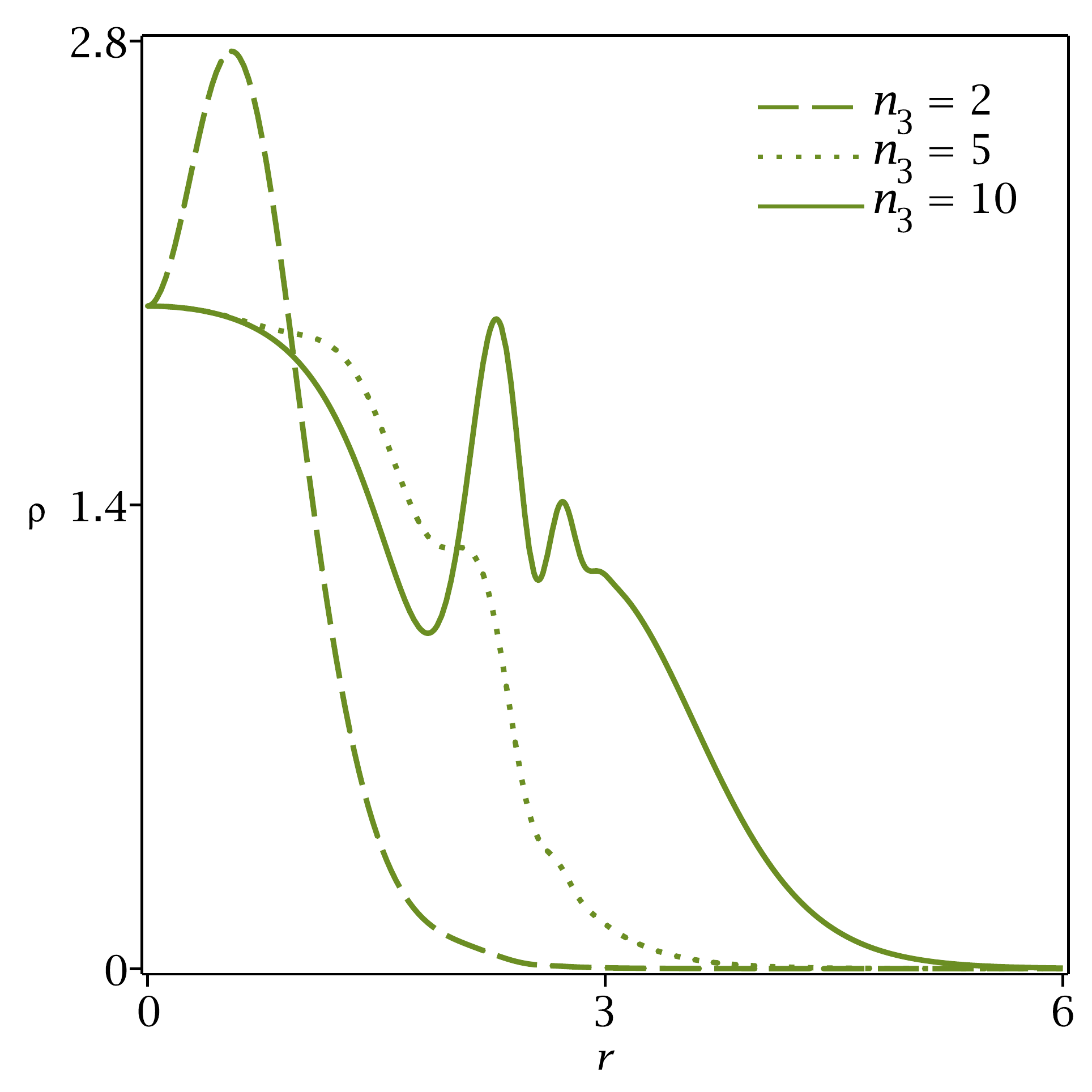}
		\caption{Solutions of~\eqref{FO} for generalized permeabilities $\mu_1=1$, $\mu_2=|\varphi_1|^2$ and $\mu_3$ given by Eq.~\eqref{ex3_5}. In the left, we depict the profiles of $g_3(r)$ (red) and $a_3(r)$ (blue). In the right, we show the corresponding energy densities $\rho_3(r)$.  Here, $n_1=1$, $n_2=10$ and  $n_3$ takes the values 2, 5 and 10.}
		\label{fig10}
	\end{figure}

\begin{figure}[h]
		\centering
		\includegraphics[width=4.8cm]{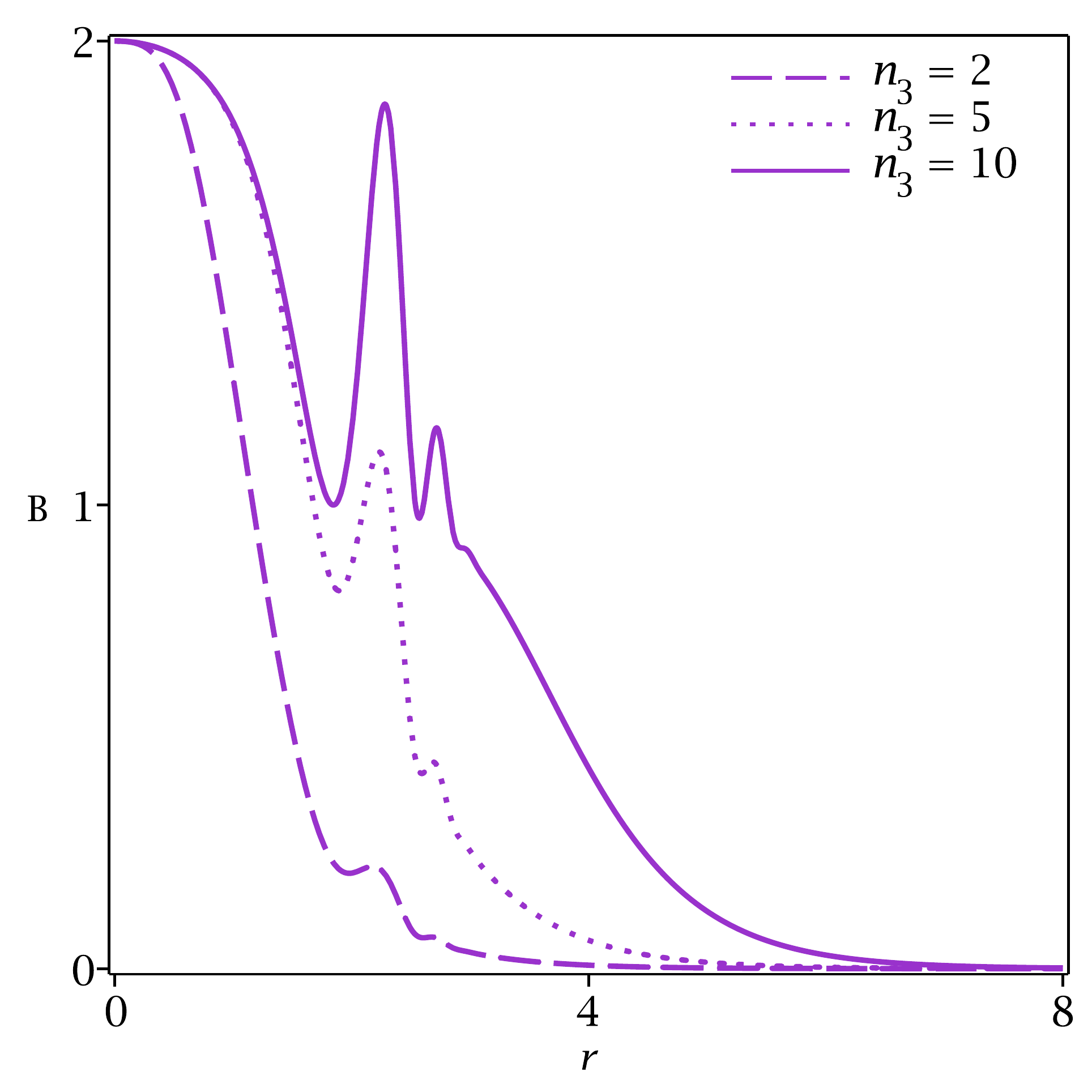}
		\caption{Magnetic inductions $B_3(r)$ for a symmetric BPS solution with permeabilities $\mu_1=1$, $\mu_2=|\varphi_1|^2$ and $\mu_3$ given by Eq.~\eqref{ex3_5}. Results are shown for three values of $n_3$, namely $n_3=2$, $n_3=5$  and $n_3=10$. }
		\label{ex4ind}
	\end{figure}

In Fig.~\ref{fig10}, we show the fields of the third sector, and the associated energy density $\rho_3$ for three values of $n_3$. In Fig.~\ref{ex4ind}, the corresponding magnetic inductions can be seen. Interestingly, this choice of $\mu_3$ results in a model whose qualitative features for fixed choices of $n_1$ and $n_2$ are strongly affected by the topological charge $n_3$ of the third sector. $B_3$ is peaked at the origin, and presents local maxima and minima generated by the oscillatory character of $\mu_3$. The number of critical points does not depend on $n_3$, but different values for this winding number produce notable distinctions in the graphs of $B_3$, as well as in the solutions, which also present internal structure. The effect of the topological charges in the graph of $\rho_3$ is even more significant, as the qualitative features of this function, as seen in the three graphs of $\rho_3(r)$ depicted in Fig.~\eqref{fig10} present very different profiles, owed solely to the change of $n_3$. This indicates a more subtle interplay between the topological charges of the three sectors in this model, all of which affect the form of the energy density.

\section{Discussion}\label{end}

Vortices and the closely related vortex-strings appear throughout many areas of physics, and gauge symmetries are used to deal with redundancies in various contexts. This versatility makes it advantageous to consider generalized models that are simple, but also highly adaptable to different contexts. In this paper, we have studied in detail a class of models defined by the introduction of generalized permeabilities in MH-like theories with enhanced symmetry. Under suitable assumptions these permeabilities may be seen as deformation functions mapping the standard theory into its modification. The effects of this deformation have been studied, often in comparison with the more well known GL and MH theories. We have seen how the presence of vortices of different kinds in a region may lead to interesting new features.

A simple class of models, with a potential motivated in connection to the GL theory, was proposed. This potential leads to a Bogomol'nyi bound and corresponding first order equations in the critical coupling. Although our examples have been worked out in this context, we have developed our theoretical considerations under milder assumptions, and have also pointed out, where appropriate, how these results may be generalized to fit the specific requirements of a given problem, as seems suitable because of the aforementioned ubiquity of vortices.

The examples discussed here are still very far from exhausting the possibilities engendered by extension of symmetry in vortex theories. Some interesting modifications include the addition of one or more Chern-Simons forms in the Lagrangian, the generalization of our results to a curved manifold and the extension of symmetry to include non-abelian groups. A further perspective comes from recent results in condensed matter, which appeared in the study of graphene, arranged to engender two \cite{GG} and three \cite{GGG1,GGG2} layers, in the last case leading to superconductivity of unconventional nature. We believe one may attempt to model vortices in multi-layer graphene by considering systems similar to those investigated in the present work. One may, for instance, consider distinct graphene layers, doping each layer with distinct percentage of Boron and Nitrogen atoms; see, e.g., \cite{Nano1,Nano2} and references therein. Also, it is possible to add an imposition of the form $A_{k}^{\{a\}}=A_{k}^{\{b\}}$, for $a\neq b$ in our equations (i.e., considering two or three scalar fields coupled to the same gauge field), at least in some bilayer and trilayer conformations. Consideration of those variations of our models appears worthwhile. Another possibility lies in dropping the assumption of additivity in the potential, which could then take a more general form. One may also consider scalar fields that are charged under multiple \emag subgroups, as done in~\cite{Impurities,ADI} for a $\rm{U(1)}\times\rm{U(1)}$ gauge theory. Some of these modifications would presumably result in a complicated coupling of the equations of motion, but their investigation could possibly be useful for modeling in a broad range of scenarios, including emergent phenomena in condensed matter physics. One of the objectives of this paper is that the discussion presented here may serve as both a motivator and a stepping stone in the investigation of extensions such as the ones discussed in this paragraph. 

\acknowledgements{This work is supported by the Brazilian agencies Coordena\c{c}\~ao de Aperfei\c{c}oamento de Pessoal de N\'ivel Superior (CAPES), grant No 88887.485504/2020-00 (MAL), Conselho Nacional de Desenvolvimento Cient\'ifico e Tecnol\'ogico (CNPq), grants No. 404913/2018-0 (DB) and No. 303469/2019-6 (DB), Federal University of Para\'\i ba (UFPB/PROPESQ/PRPG) project code PII13363-2020 and Paraiba State Research Foundation (FAPESQ-PB) grant No. 0015/2019. }

\end{document}